\documentclass[notitlepage,superscriptaddress,showpacs,nobalancelastpage,twocolumn,aps,prb,floatfix,longbibliography]{revtex4-2}
\usepackage[english]{babel}
\RequirePackage[T1]{fontenc}
\RequirePackage{times}
\usepackage{siunitx}
\usepackage[italicdiff]{physics}
\usepackage{amsfonts}
\usepackage{amsmath}
\usepackage{amssymb}
\usepackage{graphicx}
\usepackage[hidelinks]{hyperref}
\usepackage{xcolor}
\usepackage{bm}
\allowdisplaybreaks[3]

\makeatletter
\AtBeginDocument{\let\LS@rot\@undefined}
\makeatother

\begin{document}

\title{Enhancing magnonic frequency combs via geometric nonlinearity}
\author{Yushun Fang}
\affiliation{State Key Laboratory of Surface Physics and Institute for Nanoelectronic Devices and Quantum Computing, Fudan University, Shanghai 200433, China}
\affiliation{Department of Physics, Fudan University, Shanghai 200433, China}
\author{Weichao Yu}
\email{wcyu@fudan.edu.cn}
\affiliation{State Key Laboratory of Surface Physics and Institute for Nanoelectronic Devices and Quantum Computing, Fudan University, Shanghai 200433, China}
\affiliation{Zhangjiang Fudan International Innovation Center, Fudan University, Shanghai 201210, China}
\date{\today}

\begin{abstract}
Magnonic frequency combs (MFCs) generated via internal magnetic nonlinearities have exhibited rich physics beyond their optical counterparts. However, existing approaches rely on dynamic nonlinearity, which typically demands high power thresholds and stringent momentum conservation. Here we show that the geometric nonlinearity intrinsic to magnetic systems, originating from the unit-norm constraint of the magnetization vector, can serve as an independent nonlinear resource for MFC generation. Through transverse Floquet engineering, this geometric constraint converts a transverse drive into a longitudinal parametric modulation. In the low-frequency limit, the four-particle process reduces to an effective two-magnon modulation, enabling low-threshold comb generation in the linear regime, with the modulation amplitude scaling quadratically with the driving field and enhanced flatness arising from geometric harmonics. These results provide a deeper understanding of frequency combs and magnon nonlinear interactions, and offer a new theoretical foundation for enhancing MFC performance.
\end{abstract}

\maketitle

\textit{Introduction}--- Magnons, the elementary excitations of magnetic systems, feature low Joule heat dissipation and strong coupling to diverse physical systems~\cite{PhysRevLett.113.156401,shen2025cavity,yu2026electromagnetic,yuan2022quantum}, rendering them promising for classical~\cite{yu2020magnetic}, neuromorphic~\cite{grollier2020neuromorphic,zhou2021prospect}, and quantum information processing~\cite{kurebayashi2026metrics}. The precise control and measurement of magnon frequencies necessitate the development of magnonic frequency combs (MFCs), i.e., spectra consisting of discrete, equally spaced, and phase-coherent lines. To date, a rich variety of MFC generation mechanisms have exploited the internal nonlinearities of magnetic systems, including nonlinear magnon scattering in skyrmion textures~\cite{PhysRevLett.127.037202,MFC_skr_prB,liu2024low,MFC_skr_prB_THz,Liang2024Asymmetric}, stimulated three-magnon scattering under geometric confinement~\cite{zhou2021spin,Nonlinear_Doppler,Curvature_yan}, four-magnon interaction~\cite{yan_-chip_2026}, magnetic vortex dynamics~\cite{heins2026self,wang2022twisted}, magnomechanical coupling, and cavity magnonics with exceptional points~\cite{wang2024enhancement,PhysRevLett.125.237201,Liu2025,duan2025magnon,wang2026Stimulated,huang2026fractionalmagnonicfrequencycombs}. These works have shown that MFCs are not merely magnetic counterparts of optical frequency combs, but possess substantially richer physics arising from the unique nonlinear landscape of magnetic systems. Nevertheless, these approaches share intrinsic bottlenecks rooted in their reliance on \textit{dynamic nonlinearity}. For instance, three-magnon processes typically require specific momentum conservation conditions, often necessitating topological textures or geometric confinement~\cite{wang2026Stimulated}, while four-magnon processes demand high pump power thresholds~\cite{yan_-chip_2026}. 

\begin{figure*}[t]
    \centering
    \includegraphics[width=0.92\textwidth]{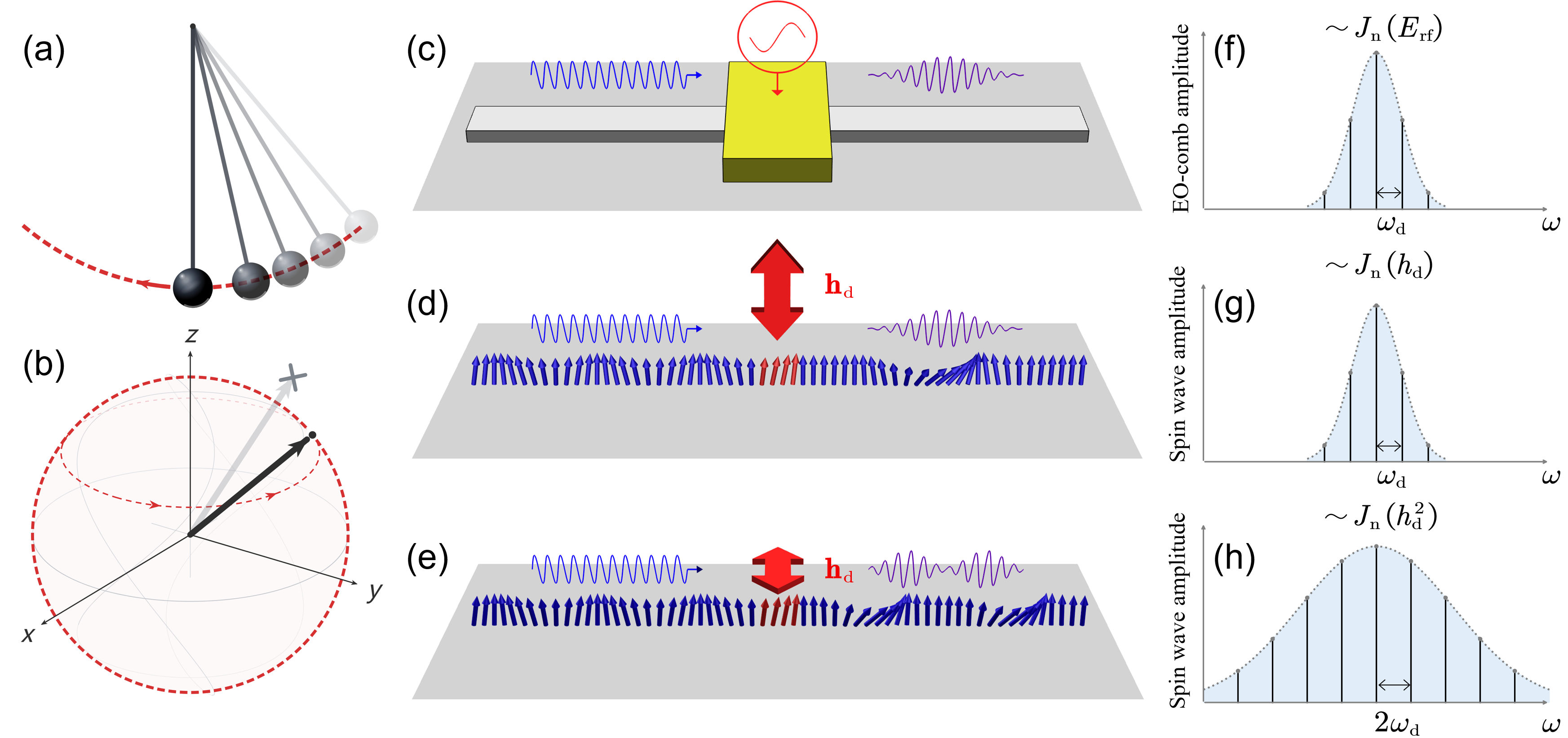}
    \caption{Geometric nonlinearity and Floquet driving. 
(a),(b) Constraint-induced geometric nonlinearity: a rigid-rod pendulum confined to a circular arc (red dashed curve) and a classical magnetic moment confined to the Bloch sphere by $|\mathbf{m}|=1$ (red dashed surface). 
(c)--(e) Conventional optical, longitudinal magnonic, and transverse magnonic Floquet driving; in (e), the unit-norm constraint converts the transverse drive into an effective longitudinal modulation. 
(f)--(h) Corresponding sideband intensities, whose amplitudes are proportional to Bessel functions $J_n$ ($n$-th order): optical $\sim J_n(E_{\mathrm{rf}})$; longitudinal $\sim J_n(h_{\mathrm{d}})$ with spacing $\omega_{\mathrm{d}}$; transverse geometric $\sim J_n(h_{\mathrm{d}}^2)$ with spacing $2\omega_{\mathrm{d}}$.}
    \label{fig:fig1}
\end{figure*}

Beyond dynamic nonlinearity, magnetic systems possess another form of nonlinearity that has received considerably less attention, namely, \textit{geometric nonlinearity} (also known as \textit{kinematic nonlinearity}), dictated by the strict unit-norm constraint of the magnetization vector, $|\mathbf{m}|=1$ \cite{Dyson1956General,Dyson1956Thermodynamic}. Figure~\ref{fig:fig1}(a) illustrates a classical-mechanical example of geometric nonlinearity, where a rigid-rod pendulum is confined to a circular arc, providing geometric nonlinearity independently of dynamical interactions such as a nonuniform gravitational field. Similarly, Fig.~\ref{fig:fig1}(b) shows the geometric nonlinearity in a magnetic system, where the magnetic moment is constrained to the Bloch sphere. This geometric constraint intrinsically correlates the transverse and longitudinal degrees of freedom and constitutes the mathematical root from which all multi-magnon scattering vertices originate via the Holstein-Primakoff expansion \cite{HolsteinPrimakoff1940,Dyson1956General}. In conventional treatments, geometric and dynamic nonlinearities are intimately intertwined: the geometric constraint provides the algebraic structure, while material-specific interaction tensors supply the dynamical coupling that drives actual scattering events. As a result, the geometric nonlinearity has not been recognized as an independent nonlinear resource, and its standalone contribution to MFC generation remains largely unexplored. This motivates a fundamental question: can the geometric nonlinearity alone, without invoking dynamic scattering processes, provide a sufficiently strong nonlinearity to generate and enhance MFCs? Answering this question is essential for deepening our understanding of the full nonlinear landscape of magnetic systems and for identifying new routes toward improved MFC performance.

In this work, we propose to address this question through transverse Floquet engineering, and our theoretical analysis predicts an affirmative answer. In optical systems, Floquet engineering uses external periodic driving to modulate the system Hamiltonian, as demonstrated in electro-optic frequency combs via the Pockels effect~\cite{zhang2019broadband,hwang2026electro,zhang2023power,hu2022high,song2026universal}. The magnetic analogue is longitudinal Floquet driving~\cite{PhysRevLett.131.243601,ye2025magnetostrictive,wang2025mechanically,xiong2023magnonic,PhysRevA.107.053708}, where an oscillating field parallel to the static bias directly modulates the magnon frequency. Here, to probe the geometric nonlinearity, we introduce a different degree of freedom, namely, \textit{transverse} Floquet driving. By applying the drive perpendicular to the static field, the unit-norm constraint converts the transverse oscillation into an effective longitudinal parametric modulation, a mechanism with no direct counterpart in scalar optical systems. We show theoretically that this purely geometric mechanism is sufficient to generate MFCs where the seed spin wave operates strictly in the linear regime, with enhanced spectral flatness arising from naturally emergent higher-order geometric harmonics, and governed by strict angular momentum selection rules.

\textit{Theoretical framework}--- Figures~\ref{fig:fig1}(c)--(e) show the three driving configurations, with the corresponding spectra in Figs.~\ref{fig:fig1}(f)--(h). For a longitudinal drive $\mathbf{h}_{\mathrm{d}} = h_{\mathrm{d}}\cos(\omega_{\mathrm{d}} t)\hat{\mathbf{z}}$ parallel to the static field $\mathbf{H}_0$, the Zeeman energy is directly modulated, yielding a linear frequency shift $\omega_k^{\parallel}(t) = \omega_k + \gamma h_{\mathrm{d}}\cos(\omega_{\mathrm{d}} t)$, with $\omega_k = \gamma H_0 + \gamma A k^2$ the unperturbed magnon frequency. For a transverse drive $\mathbf{h}_{\mathrm{d}} = h_{\mathrm{d}}\cos(\omega_{\mathrm{d}} t)\hat{\mathbf{x}}$, the drive does not directly alter the longitudinal field, but the unit-norm constraint forces $m_z = \sqrt{1 - m_x^2} \approx 1 - \frac{1}{2}(h_{\mathrm{d}}/H_0)^2\cos^2(\omega_{\mathrm{d}} t)$, which contains a static reduction and a $2\omega_{\mathrm{d}}$ component. This geometric nonlinearity converts the transverse oscillation into an effective longitudinal parametric modulation. Combining this with the Zeeman energy of the drive, $h_{\mathrm{d}}\cos(\omega_{\mathrm{d}} t)m_x$, the precession frequency can be expressed as

\begin{equation}
\omega_k^{\perp}(t) = \tilde{\omega}_k + V\cos(2\omega_{\mathrm{d}} t), \qquad V = \frac{\gamma h_{\mathrm{d}}^2}{4H_0}, \label{eq.classical_modulation}
\end{equation}
where $\tilde{\omega}_k = \omega_k + \gamma h_{\mathrm{d}}^2/(4H_0)$ is the renormalized magnon frequency shifted by the time-averaged geometric constraint. This frequency-doubling effect is known in classical nonlinear magnetization theory~\cite{gurevich1996magnetization}. Here we promote it to the core mechanism of transverse Floquet engineering for MFC generation. The resulting sideband spacing is therefore $2\omega_{\mathrm{d}}$ [Fig.~\ref{fig:fig1}(h)], in contrast to the $\omega_{\mathrm{d}}$ spacing of longitudinal driving [Fig.~\ref{fig:fig1}(g)].

The transverse modulation amplitude $V = \gamma h_{\mathrm{d}}^2/(4H_0)$ scales quadratically with $h_{\mathrm{d}}$, unlike the linear longitudinal shift $\gamma h_{\mathrm{d}}$ (see Supplementary Materials (SM)~\cite{SM} for the macrospin picture and the local-comoving-frame derivation).

We now derive Eq.~\eqref{eq.classical_modulation} quantum mechanically to reveal its microscopic origin. The classical Hamiltonian is

\begin{equation}
\mathcal{H} = M_s \int \left[-\mu_0 \mathbf{m} \cdot \left(\mathbf{H}_0 + \mathbf{h}_{\mathrm{d}}\right) + A\left(\boldsymbol{\nabla}\mathbf{m}\right)^2\right] d^3r, \label{eq.Hamil}
\end{equation}
where $M_s$ is the saturation magnetization, $\mu_0$ the vacuum permeability, $\mathbf{m}$ the normalized magnetization, $\mathbf{H}_0$ the static field, $\mathbf{h}_{\mathrm{d}}$ the driving field, and $A$ the exchange coefficient.

We apply the Holstein-Primakoff (HP) transformation to the spin waves and quantize the transverse drive into a photon mode $\hat{c}$~\cite{rezende2020fundamentals}. Since $\omega_{\mathrm{d}} \ll \omega_k$, higher-order HP terms must be retained~\cite{zheng2023tutorial,Dyson1956General}. The unit-norm constraint manifests in $\hat{S}_z = S - \hat{a}^\dagger \hat{a}$ (with $S$ the total spin length and $\hat{a}$ the magnon annihilation operator): transverse magnons deplete the longitudinal spin \cite{HolsteinPrimakoff1940,Dyson1956General}. The interaction term arises from the second-order expansion of $S_x$ coupled to $\hat{c}$, yielding

\begin{equation}
\hat{H}_{\mathrm{int}}^{(2)} = g_2 \sum_{\mathbf{k}} \left[ \left(\hat{c}^\dagger + \hat{c} \right) \hat{a}_{\mathbf{k}}^\dagger \hat{a}_{\mathbf{k}} \hat{a}_0 + \mathrm{h.c.} \right], \label{eq.four_particle}
\end{equation}
where $g_2$ is the coupling strength, $\hat{a}_{\mathbf{k}}$ the magnon operator at wave vector $\mathbf{k}$, and $\hat{a}_0$ the $\mathbf{k}=0$ mode operator. This term is a photon-assisted three-magnon operator product, in which the $\mathbf{k}=0$ mode, arising from the forced oscillation of the magnetization driven by the modulation, modulates the density of the $\mathbf{k}$ mode and represents the geometric constraint microscopically~\cite{Dyson1956General} (see SM~\cite{SM} for the expansion of higher-order geometric harmonics).

Since $\omega_{\mathrm{d}} \ll \omega_k$, the drive only excites the $\mathbf{k}=0$ mode. Replacing $\hat{a}_0$ and $\hat{c}$ by their expectation values $\langle\hat{a}_0\rangle \propto (h_{\mathrm{d}}/H_0)e^{-i\omega_{\mathrm{d}} t}$ and $\langle\hat{c}\rangle \propto h_{\mathrm{d}} e^{-i\omega_{\mathrm{d}} t}$ collapses the interaction into a two-magnon parametric Hamiltonian at $2\omega_{\mathrm{d}}$:

\begin{equation}
\hat{H}_{\mathrm{eff}} = V\cos(2\omega_{\mathrm{d}} t)\,\hat{a}_{\mathbf{k}}^\dagger\hat{a}_{\mathbf{k}}, \label{eq.effective}
\end{equation}
with $V$ identical to the classical result. The quantum derivation shows that, upon applying a mean-field approximation to two of the four operators, the four-operator interaction reduces to an effective two-magnon term that is strictly diagonal. Importantly, the resulting nonlinear interaction is determined solely by the external transverse drive, rather than by the injected spin-wave signal. The seed spin wave can therefore remain in the linear regime while the comb is generated, circumventing the power thresholds associated with dynamic nonlinearities. Hereafter, we omit the subscript $\mathbf{k}$.

To solve for the steady-state spectrum, we include the injected signal $h_0$ at $\omega_0$ (the external signal frequency) and Gilbert damping $\alpha$. Defining $\delta = \tilde{\omega}_k - \omega_0$, the Heisenberg equation reads

\begin{equation}
\frac{d\hat{a}}{dt} = -i\left[\left(\delta - i\alpha\tilde{\omega}_k\right)\hat{a} + \frac{\gamma h_0}{2} + V\cos(2\omega_{\mathrm{d}} t)\,\hat{a}\right], \label{eq.heisenberg}
\end{equation}
where $-i\alpha\tilde{\omega}_k$ accounts for Gilbert damping. The drive induces a phase factor $e^{-i\Phi(t)}$ with $\Phi(t) = \frac{V}{2\omega_{\mathrm{d}}}\sin(2\omega_{\mathrm{d}} t)$. The Jacobi-Anger expansion discretizes the spectrum into sidebands spaced by $2\omega_{\mathrm{d}}$, mapping onto a dissipative Wannier-Stark ladder~\cite{GRIFONI1998229} (see SM~\cite{SM}). The Floquet Green's function $G^{\mathrm{F}} \equiv (H_{\mathrm{F}} + i\alpha\tilde{\omega}_k I)^{-1}$ gives the $l$-th sideband response:

\begin{equation}
G^{\mathrm{F}}_{l0} = \sum_{n=-\infty}^{\infty}\frac{J_{n-l}(\beta)\,J_n(\beta)}{\delta + 2n\omega_{\mathrm{d}} - i\alpha\tilde{\omega}_k}, \label{eq.GF}
\end{equation}
where $J_n$ is the $n$-th Bessel function, and $\beta = V/(2\omega_{\mathrm{d}}) = \gamma h_{\mathrm{d}}^2/(8H_0\omega_{\mathrm{d}})$ is the modulation index. Therefore, we reach the analytical solution for the amplitude of $l$-th physical sideband: $\mathcal{A}_l = -(\gamma h_0/2)\, G^{\mathrm{F}}_{l0}$.

\textit{Numerical verification}--- We perform micromagnetic simulations with COMSOL~\cite{xu_frequency_2025,Wang2026,comsol,Zhang2023} solving the Landau-Lifshitz-Gilbert equation:
\begin{equation}
    \frac{d\mathbf{m}}{dt} = -\gamma\mathbf{m} \times \mathbf{H}_{\mathrm{eff}} + \alpha \mathbf{m} \times \frac{d\mathbf{m}}{dt}, \label{eq.LLG}
\end{equation}
where $\mathbf{H}_{\mathrm{eff}} = -\frac{1}{\mu_0 M_s} \frac{\delta \mathcal{H}}{\delta \mathbf{m}}$ is the effective field derived from the Hamiltonian in Eq.~\eqref{eq.Hamil}. We simulate a 1D spin chain with $\mathbf{H}_0 = H_0\hat{\mathbf{z}}$, $H_0 = 8.48 \times 10^4$~A/m (Kittel frequency 3~GHz). A 7~GHz spin wave ($\lambda = 106.85$~nm) passes through a $\lambda/2$ modulation region, where a transverse microwave field of frequency $\omega_{\mathrm{d}}$ and amplitude $h_{\mathrm{d}}$ is applied. Both ends have $8\lambda$ absorbing regions ($\alpha=0.2$) to prevent reflections (bulk $\alpha=0.001$). The spin-wave amplitude is kept small within the linear regime, and dipolar interactions are neglected. The transmitted signal is Fourier transformed. YIG parameters~\cite{lan_spin-wave_2015}: $A = 0.328 \times 10^{-10}~\mathrm{A\cdot m}$, $\gamma = 2.21 \times 10^{5}~\mathrm{rad\,s^{-1}/(A/m)}$, $M_s = 0.194 \times 10^{6}~\mathrm{A/m}$.

\begin{figure}[t]
    \centering
    \includegraphics[width=\columnwidth]{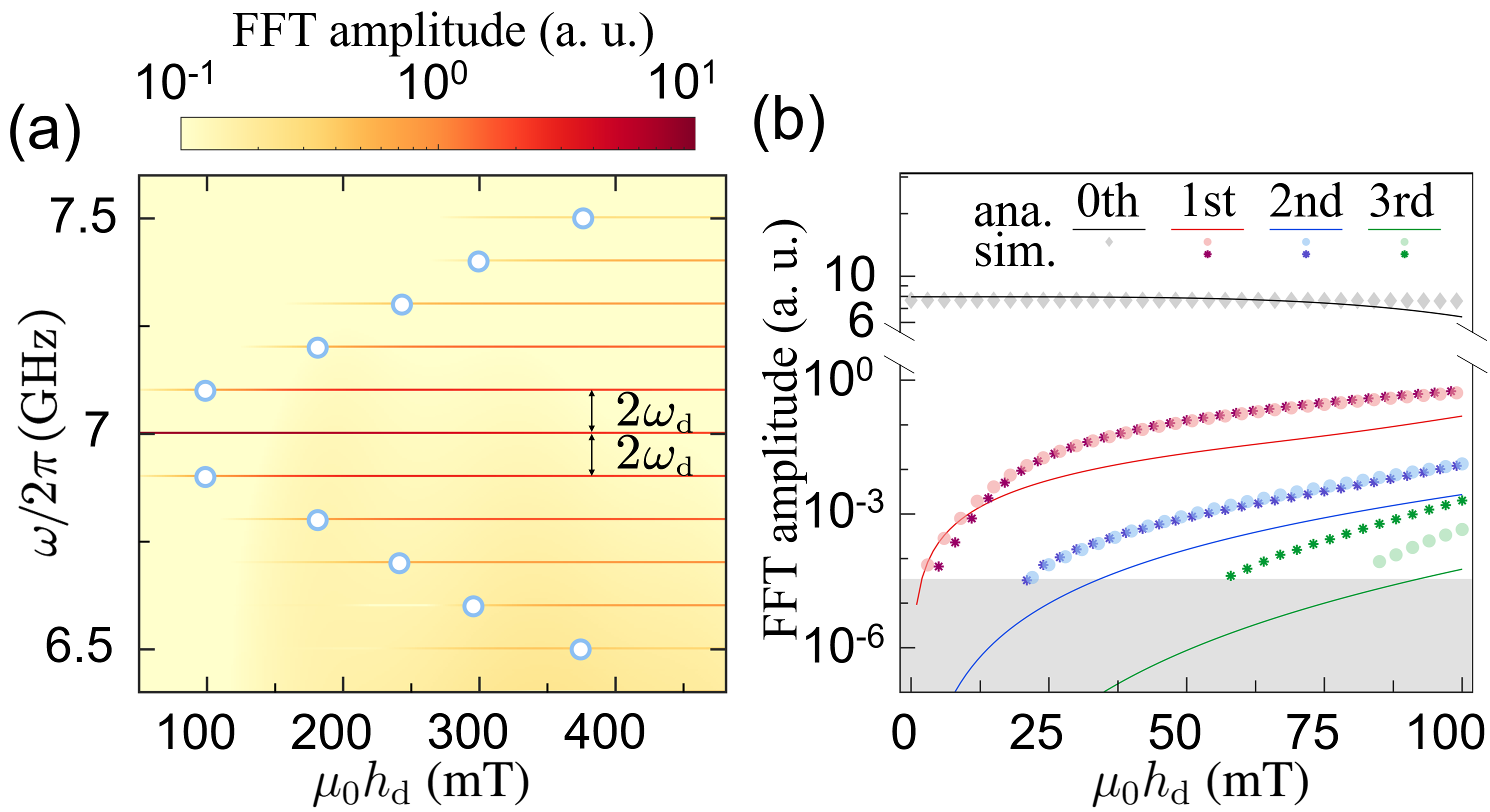}
    \caption{Comparison between theoretical predictions and numerical simulations for the transverse Floquet driving scheme. (a) Evolution of the MFC with increasing driving amplitude. (b) Intensity comparison for the first three comb teeth.}
    \label{fig:fig2}
\end{figure}

Figure~\ref{fig:fig2}(a) shows the growth of the frequency comb with increasing driving amplitude, where blue circles mark the theoretically predicted $h_{\mathrm{d}}$ at which each higher-order tooth first reaches $1/50$ of the central (0th-order) tooth intensity. The intensities of the first three comb teeth versus $h_{\mathrm{d}}$ are compared with analytical results in Fig.~\ref{fig:fig2}(b) (solid curves: theory; dots: simulation; gray shading: noise floor). At low driving fields, simulation agrees well with theory. As $h_{\mathrm{d}}$ increases, higher-order teeth grow faster than predicted, because the analytical model truncates the HP expansion at the lowest nonlinear order (small-angle precession, $h_{\mathrm{d}} \ll H_0$). Large-angle precession activates higher-order geometric harmonics and multi-magnon channels, providing additional pumping and enhancing the effective nonlinearity (see SM~\cite{SM} for the expansion of higher-order geometric harmonics). Within the simulated parameter range, the system remains in a stable Floquet state and does not show signatures of chaos or modulation instability~\cite{yan_-chip_2026} (see SM~\cite{SM} for additional numerical results, including Floquet states in the dispersion relation and Mumax3 simulations in thin films).

\textit{Sum- and difference-frequency generation by multiple polarized drives}--- We now extend to circularly polarized drives. Helicity $\sigma = \pm 1$: $\sigma = +1$ for $\mathbf{h}_{\mathrm{d}} = h_{\mathrm{d}}[\cos(\omega_{\mathrm{d}} t)\,\hat{\mathbf{x}} + \sin(\omega_{\mathrm{d}} t)\,\hat{\mathbf{y}}]$, $\sigma = -1$ opposite (see SM~\cite{SM} for the polarization analysis). We fix a drive with $\sigma_1 = +1$ at $f_1 = 0.1$ GHz and vary the second drive.

\begin{figure}[htb]
    \centering
\includegraphics[width=\columnwidth]{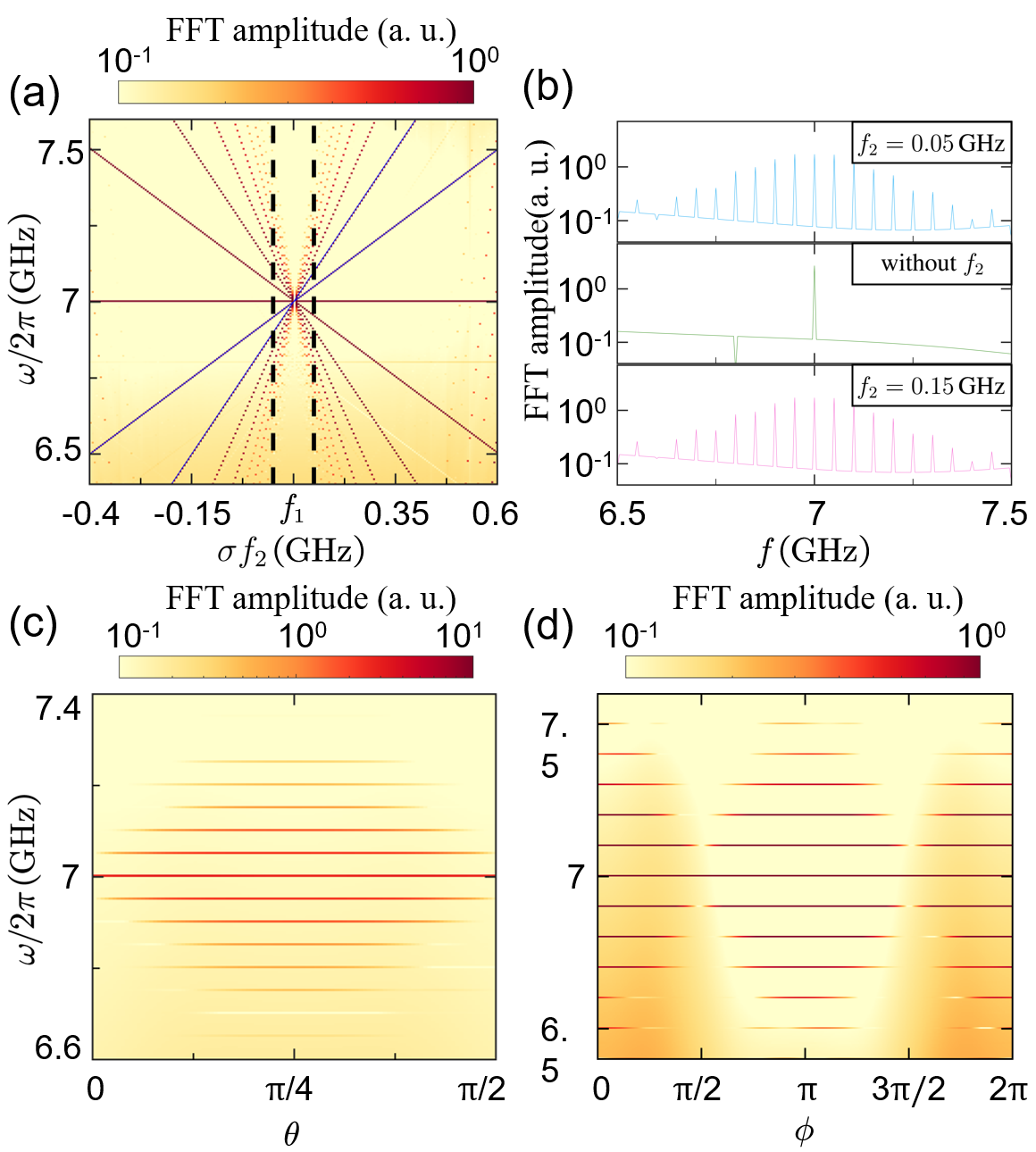}
    \caption{Floquet driving with circularly polarized fields. (a) Repetition frequency versus the second driving frequency, with opposite helicities giving the sum frequency and same helicities giving the difference frequency. (b) MFC spectra for selected configurations. (c) Comb spectrum versus $\theta$ under two drives with constant total intensity. (d) Comb spectrum versus $\phi$ under a single drive.}
    \label{fig:fig3}
\end{figure}

Figure~\ref{fig:fig3}(a) shows the repetition frequency versus the second driving frequency. Solid lines mark the calculated first- and second-order comb teeth positions. When the second drive has opposite helicity ($\sigma_2 = -\sigma_1$), the repetition frequency is the sum; when same helicity ($\sigma_2 = \sigma_1$), it is the difference. Calculated positions agree with simulations. Figure~\ref{fig:fig3}(b) shows selected spectra. No frequency comb is generated under a single circularly polarized drive. The $\mathbf{k}=0$ magnon mode is excited through second-order coupling to the photon mode and therefore inherits its frequency and helicity~\cite{gurevich1996magnetization}. For a single circularly polarized photon mode, the time-dependent phases of the $\mathbf{k}=0$-magnon operator and the corresponding photon operator cancel in the four-operator interaction, yielding a time-independent coefficient after the mean-field reduction. Consequently, no periodic frequency modulation is generated. With two photon modes of different frequencies and/or helicities, the $\mathbf{k}=0$ magnon mode excited by one photon mode can instead couple to the other, producing a periodic modulation at the sum frequency for opposite helicities or at the difference frequency for the same helicity. Linear or elliptical polarization contains both $\sigma=+1$ and $\sigma=-1$ components and thus provides the opposite-helicity channel required for comb generation (see SM~\cite{SM}).

Fixing $f_1 = 0.1$ GHz and $f_2 = 0.05$ GHz with $h_1 = h_{\mathrm{d}}\cos\theta$, $h_2 = h_{\mathrm{d}}\sin\theta$, Fig.~\ref{fig:fig3}(c) shows the comb spectrum versus $\theta$; the largest number of teeth appears when $h_1 = h_2$. Figure~\ref{fig:fig3}(d) varies $\phi$ for a single drive, $\mathbf{h}_{\mathrm{d}} = h_{\mathrm{d}}[\cos(\omega_{\mathrm{d}} t)\hat{\mathbf{x}} + \sin\phi\,\sin(\omega_{\mathrm{d}} t)\hat{\mathbf{y}}]$ ($\phi = n\pi$ linear, $(2n+1)\pi/2$ circular). The comb is most pronounced at $\phi = n\pi$ (linear polarization), which contains equal $\sigma = \pm 1$ components. Since low-frequency microwave sources are challenging to implement, this same-helicity difference-frequency channel offers a practical route to low repetition rates by superposing two high-frequency co-polarized drives with closely spaced frequencies.

\textit{Spectral flatness and benchmarking}--- We define the spectral roll-off rate $\mathcal{R}$ as~\cite{wang2026ultralow,10969078,9360806,Xiang_2023}

\begin{equation}
\mathcal{R} = \frac{P_n - P_0}{n}, \label{eq.roll_off}
\end{equation}
where $P_0$ ($P_n$) is the carrier ($n$-th tooth) power. A smaller $|\mathcal{R}|$ indicates a flatter spectrum, with $\mathcal{R} = 0$ corresponding to a perfectly flat comb. Throughout this work, we take $n = 3$ to ensure a consistent comparison. Some works use the opposite sign; all literature data are converted to this convention. We study $\mathcal{R}$ versus repetition frequency.

\begin{figure}[htbp]
    \centering
    \includegraphics[width=\columnwidth]{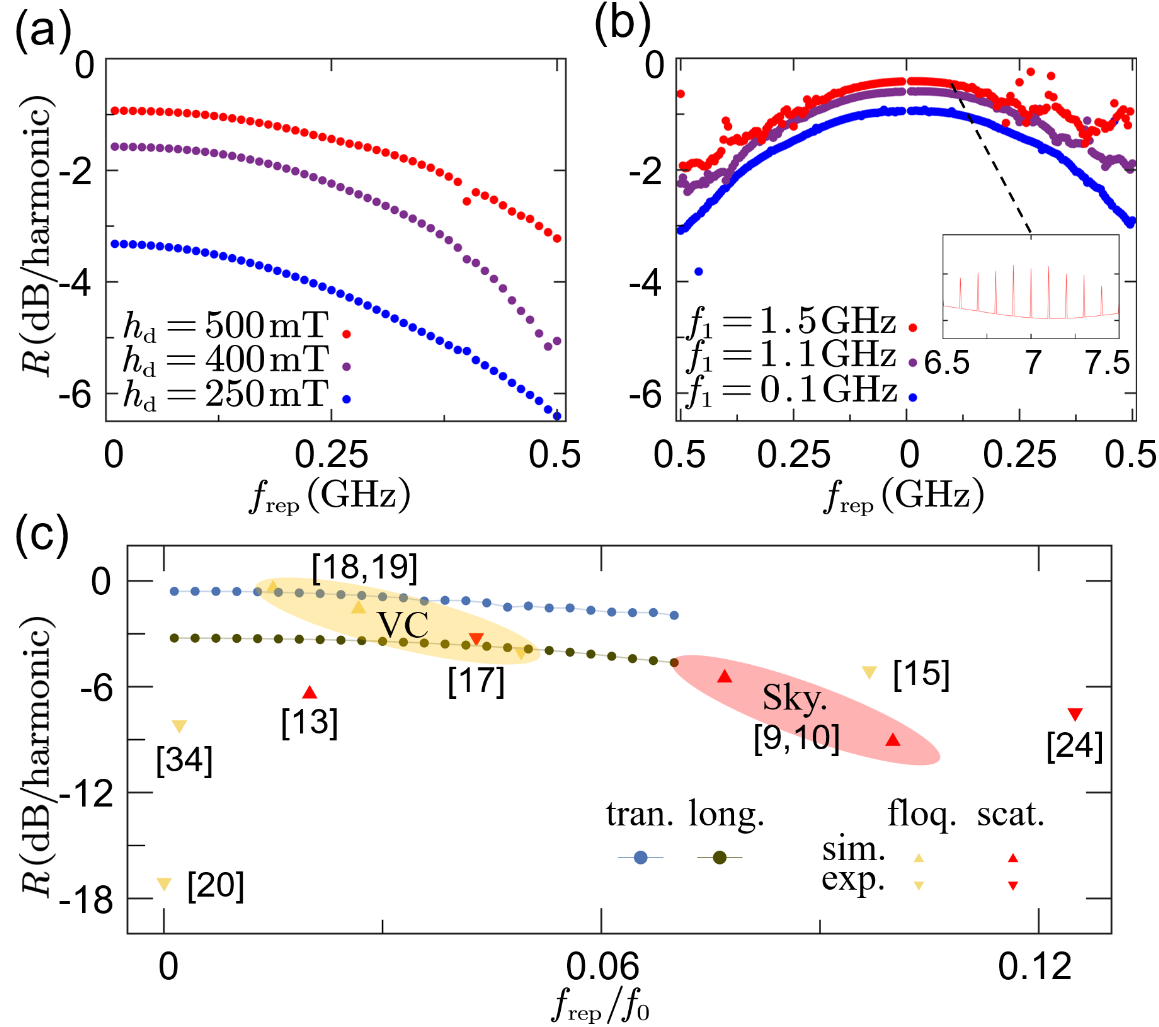}
    \caption{Comparison of spectral flatness. (a) Spectral roll-off rate $\mathcal{R}$ as a function of repetition frequency under linearly polarized driving. (b) Variation of $\mathcal{R}$ for two equal-intensity circularly polarized drives. (c) Flatness performance comparison among different MFC generation mechanisms.}
    \label{fig:fig4}
\end{figure}

Figure~\ref{fig:fig4}(a) shows $\mathcal{R}$ versus repetition frequency under linearly polarized driving at different drive amplitudes. Deviations are due to interference. A higher drive amplitude leads to a larger $\mathcal{R}$ (closer to zero), i.e., a smaller $|\mathcal{R}|$, indicating enhanced flatness. Figure~\ref{fig:fig4}(b) varies $f_1$ while sweeping the second frequency. $\mathcal{R}$ increases toward zero with $f_1$, indicating enhanced flatness. This behavior arises because a higher $f_1$ raises the overall driving frequency, so that the low-frequency approximation gradually breaks down and higher-order nonlinear interactions are activated, which flatten the spectrum.

Finally, we benchmark against other MFC schemes. In Fig.~\ref{fig:fig4}(c), dashed lines are our transverse/longitudinal simulations, triangles are literature data (converted). We normalize by $f_{\mathrm{rep}}/f_0$ and use $n=3$ for all data \cite{PhysRevLett.127.037202,MFC_skr_prB_THz,Liang2024Asymmetric,yan_-chip_2026,wang2026Stimulated,PhysRevLett.131.243601,wang2024enhancement,heins2026self,Nonlinear_Doppler,wang2022twisted}. The transverse scheme yields significantly smaller $|\mathcal{R}|$ in most cases, indicating superior flatness. This is because the inherent anharmonicity of the geometric constraint naturally introduces higher-order geometric harmonics ($4\omega_{\mathrm{d}}, 6\omega_{\mathrm{d}}$) at moderate to strong driving fields. These multi-frequency modulations fill the spectral dips characteristic of single-frequency Bessel envelopes, leading to a flatter spectrum. The Floquet driving via magnetic vortex dynamics performs comparably in some regimes, since the gyrotropic motion also involves a transverse driving component~\cite{heins2026self,wang2022twisted}. Since geometric nonlinearity is universally present in magnetic systems, how it contributes to existing MFC schemes remains an open question.

\textit{Conclusion and discussion}--- In conclusion, we have shown that the geometric nonlinearity intrinsic to magnetic systems, dictated by the unit-norm constraint of the magnetization vector, can be isolated from dynamic scattering processes and harnessed as an independent nonlinear resource for MFC generation. Transverse Floquet engineering converts a transverse drive into a longitudinal parametric modulation: the formal four-operator magnon--photon interaction collapses into a strictly diagonal two-magnon process in the low-frequency limit, decoupling the comb-generating nonlinearity from the injected signal so that the seed spin wave remains in the linear regime. The inherent anharmonicity of the spherical manifold generates higher-order geometric harmonics that fill the dips of the single-frequency Bessel envelope and yield enhanced spectral flatness. We further establish strict angular momentum selection rules, showing that the repetition frequency can be tuned by superposing drives of different polarizations. Our results suggest a shift in the paradigm of MFC generation from interaction-induced dynamic scattering to constraint-induced geometric modulation. 

\textit{Acknowledgments}--- This work was supported by Fundamental and Interdisciplinary Disciplines Breakthrough Plan of the Ministry of Education of China (Grant No. JYB2025XDXM120), National Key Research Program of China (Grant No. 2022YFA1403300), the Innovation Program for Quantum Science and Technology (Grant No. 2024ZD0300103) and National Natural Science Foundation of China (Grant No. 12574112 and 12204107).

\clearpage
\onecolumngrid        
\appendix
\setcounter{secnumdepth}{2}

\begin{center}
\textbf{\large Supplemental Materials for ``Enhancing magnonic frequency combs via geometric nonlinearity''}
\end{center}

\setcounter{equation}{0}
\setcounter{figure}{0}
\setcounter{section}{0}
\renewcommand{\theequation}{S\arabic{equation}}
\renewcommand{\thefigure}{S\arabic{figure}}
\renewcommand{\figurename}{FIG.}
\renewcommand{\tablename}{TABLE}

\section{MACROSPIN MODEL}
\label{sec:macrospin}

Let us consider a macrospin under a static magnetic field $\mathbf{H}_0$ along the $z$ axis and a linearly polarized driving field $\mathbf{h}_{\mathrm{d}}(t)$ along the $x$ axis. The frequency of the driving field $\omega_{\mathrm{d}}$ is far below the Kittel mode frequency $\omega_K=\gamma H_0$, where $\gamma$ is the gyromagnetic ratio. We assume that the motion of the macrospin can be divided into two parts: the precession around the axis $\mathbf{H}_0+\mathbf{h}_{\mathrm{d}}$, which makes a perturbation around the $z$ axis. In that case, the precession can be taken as an adiabatic process compared with the variation of the axis. We further assume that the precession frequency is equal to $\gamma \sqrt{H_{0}^{2}+\left( h_{\mathrm{d}}\left( t \right) \right) ^2}$. In the case of $h_{\mathrm{d}} \ll H_0$, through Taylor expansion, we derive
\begin{equation}
    \omega_{k=0}^\perp(t) =\gamma \sqrt{H_{0}^{2}+\left( h_{\mathrm{d}}\left( t \right) \right) ^2}\approx \gamma H_0\left( 1+\frac{1}{2}\frac{\left( h_{\mathrm{d}}\left( t \right) \right) ^2}{H_{0}^{2}} \right) = \omega_K + \frac{\gamma h_{\mathrm{d}}^2(t)}{2H_0}.
\end{equation}
For the linearly polarized field $h_{\mathrm{d}}(t)=h_{\mathrm{d}}\cos \left( \omega_{\mathrm{d}} t \right)$, Eq.~(S1) becomes
\begin{equation}
    \omega_{k=0}^\perp(t) = \tilde{\omega}_{k=0} + V\cos(2\omega_{\mathrm{d}} t), \label{variation kittel}
\end{equation}
where $\tilde{\omega}_{k=0} = \omega_K + \frac{\gamma h_{\mathrm{d}}^2}{4H_0}$ is the renormalized precession frequency and $V = \frac{\gamma h_{\mathrm{d}}^2}{4H_0}$ is the modulation amplitude. Equation~\eqref{variation kittel} shows that the precession frequency varies near the Kittel mode frequency $\omega_K$. This frequency doubling originates from the geometric nonlinearity dictated by the unit-norm constraint of the magnetization vector. To further investigate the motion, we divide the magnetization into two parts, $\mathbf{m}=\mathbf{m}_0+\delta \mathbf{m}$. The component $\mathbf{m}_0$ can be viewed as slowly varying compared with the magnon precession, since it rotates at the frequency $\omega_{\mathrm{d}}$. The precession part is written as $\delta m_x \left( t \right)=\mathrm{Re}\left\{ m_{x0}e^{i\int_{0}^{t} \omega_{k=0}^\perp\left( t^{\prime } \right) dt^{\prime } } \right\}$, $\delta m_y \left( t \right)=\mathrm{Re}\left\{ m_{y0}e^{i\int_{0}^{t} \omega_{k=0}^\perp\left( t^{\prime } \right) dt^{\prime } } \right\}$. Applying the Jacobi-Anger expansion
$ e^{iz\sin \theta}=\sum_{n=-\infty}^{\infty}{J_n\left( z \right) e^{in\theta}}$, where $J_n$ denotes the $n$-th Bessel function of the first kind, we have
\begin{align}
    \delta m_x\left( t \right)&= \mathrm{Re}\left\{ m_{x0}e^{i\tilde{\omega}_{k=0} t}e^{i\beta\sin \left( 2\omega _{\mathrm{d}}t \right)} \right\}  \notag
\\
&=\mathrm{Re}\left\{ m_{x0}e^{i\tilde{\omega}_{k=0} t}\sum_{n=-\infty}^{\infty}{J_n\left( \beta \right) e^{in2\omega _{\mathrm{d}}t}} \right\}.    \label{Jacobi Anger expansion}
\end{align}
Here the modulation index is defined as $\beta = V/(2\omega_{\mathrm{d}}) = \gamma h_{\mathrm{d}}^2/(8H_0\omega_{\mathrm{d}})$. In the same way, $\delta m_y\left( t \right)$ can also be written in this form. Equation~\eqref{Jacobi Anger expansion} shows that the magnetization precession has a central frequency $\tilde{\omega}_{k=0}$ and sidebands at $n2\omega _{\mathrm{d}}$. The amplitude of the $n$-th sideband is $J_n\left( \beta \right)$.

In the following sections, $\omega_k = \gamma H_0 + \gamma A k^2$ denotes the magnon frequency including the exchange contribution, where $A$ is the exchange coefficient; the macrospin precession frequency discussed here corresponds to the $k=0$ limit, $\omega_{k=0}^\perp(t) = \tilde{\omega}_{k=0} + V\cos(2\omega_{\mathrm{d}} t)$.

\begin{figure*}[htb]
    \centering
    \includegraphics[width=0.7\textwidth]{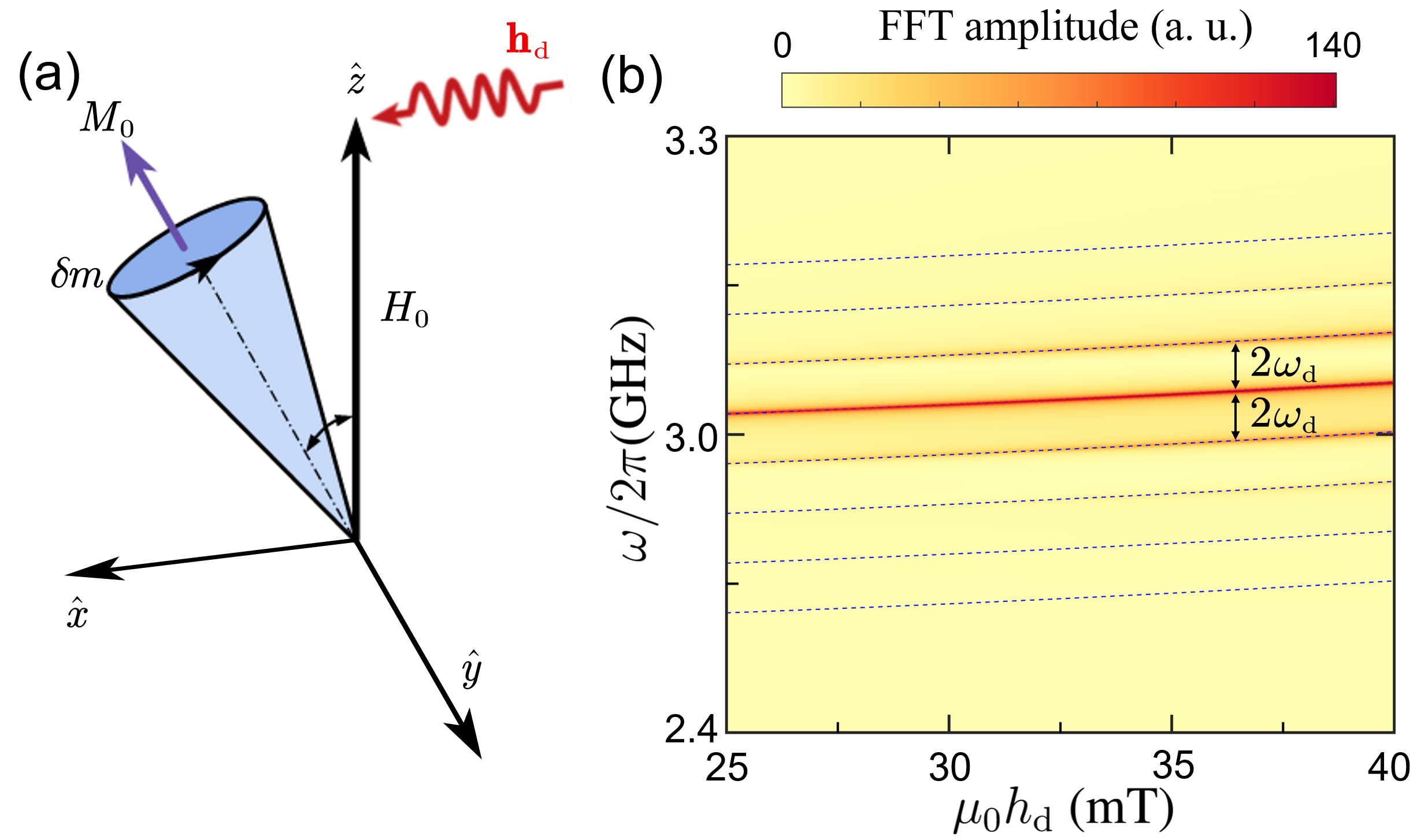}
    \caption{Macrospin model and Floquet magnonic frequency comb. (a) Schematic of the macrospin dynamics under a static field and a transverse drive. (b) MFC generation in the macrospin model, showing an increase in the number of comb teeth and the center frequency with the driving amplitude $h_{\mathrm{d}}$.}
    \label{fig:SM_fig1}
\end{figure*}

\section{ONE-DIMENSIONAL MODEL}
\label{sec:1D}

\noindent\textit{Note on the theoretical frameworks:} In this section, we present two complementary theoretical approaches to describe the transverse Floquet engineering. The first approach (Sec.~\ref{sec:classical}) employs a classical treatment of the electromagnetic field within a local comoving frame, providing an intuitive picture of the geometric nonlinearity dictated by the unit-norm constraint~\cite{Dyson1956General,Dyson1956Thermodynamic}. The second approach (Sec.~\ref{sec:quantum}) treats the electromagnetic field quantum mechanically in the laboratory frame, systematically deriving the multi-magnon interaction Hamiltonian via the Holstein-Primakoff expansion~\cite{HolsteinPrimakoff1940,Dyson1956General}. Both approaches yield identical results in the mean-field and weak-field limits, confirming the validity of the effective $2\omega_{\mathrm{d}}$ Floquet driving across these regimes.

\subsection{Theoretical framework I: Classical treatment of the electromagnetic field}
\label{sec:classical}

We first treat the electromagnetic field classically. The magnetization dynamics under the applied magnetic field is governed by the Landau-Lifshitz-Gilbert (LLG) equation:
\begin{equation}
    \frac{d\mathbf{m}}{dt} = -\gamma\mathbf{m} \times \mathbf{H}_{\mathrm{eff}} + \alpha \mathbf{m} \times \frac{d\mathbf{m}}{dt},
\end{equation}
where $\mathbf{m}$ is the normalized magnetization, $\gamma$ is the gyromagnetic ratio, $\alpha$ is the dimensionless Gilbert damping constant, and $\mathbf{H}_{\mathrm{eff}}$ denotes the effective magnetic field. The total energy of the system includes the Zeeman energy and the exchange interaction energy, with the corresponding Hamiltonian given by
\begin{equation}
\mathcal{H} = M_s \int \left[ -\mu_0 \mathbf{m} \cdot (\mathbf{H}_0 + \mathbf{h}_{\mathrm{d}}) + A (\boldsymbol{\nabla} \mathbf{m})^2 \right] d^3r  \label{eq.HamilSM}
\end{equation}
where $\mu_0$ is the vacuum permeability, $M_s$ is the saturation magnetization, $\mathbf{H}_0$ is the static magnetic field, $\mathbf{h}_{\mathrm{d}}$ is the time-dependent driving field, and $A$ is the exchange stiffness. Since $\omega_{\mathrm{d}} \ll \omega_K$, the magnetization motion naturally separates into two components with vastly different frequencies, consistent with the macrospin model discussed in the preceding section. Within the adiabatic approximation, the magnon dynamics can be treated as precession around a slowly evolving ground state. We therefore introduce a local coordinate system that comoves with the instantaneous orientation of the ground-state magnetization $\mathbf{M}_0$, with the local $\mathbf{z}$-axis defined along $\mathbf{M}_0$. The transformation between the laboratory frame and this rotating local frame is given by

\begin{align}
\begin{pmatrix} m_x \\ m_y \\ m_z \end{pmatrix}
&=
\begin{pmatrix} 0 & \cos\theta & \sin\theta \\ -1 & 0 & 0 \\ 0 & -\sin\theta & \cos\theta \end{pmatrix}
\begin{pmatrix} m_1 \\ m_2 \\ m_3 \end{pmatrix}.    \label{eq.trans}
\end{align}

The angle $\theta = \arctan \left( h_{\mathrm{d}} \cos(\omega_{\mathrm{d}} t) / H_0 \right)$ denotes the tilt angle between $\mathbf{M}_0$ and the $\mathbf{z}$-axis of the laboratory frame. In the local frame, we apply the Holstein-Primakoff transformation~\cite{HolsteinPrimakoff1940} to the magnetization components $m_1$, $m_2$, and $m_3$, retaining only the linear terms:
\begin{align}
m_1 &\approx \frac{1}{\sqrt{2S}} \left( \hat{a} + \hat{a}^\dagger \right), \notag \\
m_2 &\approx -\frac{i}{\sqrt{2S}} \left( \hat{a} - \hat{a}^\dagger \right), \notag \\
m_3 &= 1 - \frac{\hat{a}^\dagger \hat{a}}{S}, \label{eq:HP}
\end{align}
where $\hat{a}$ ($\hat{a}^\dagger$) denotes the magnon annihilation (creation) operator, and $S$ is the spin quantum number. In the instantaneous local frame, the linear terms vanish due to the equilibrium condition, and the lowest-order Zeeman energy contribution is proportional to $m_3$, justifying the retention of only the $\hat{a}^\dagger\hat{a}$ term. Substituting Eqs.~\eqref{eq.trans} and \eqref{eq:HP} into the Hamiltonian \eqref{eq.HamilSM} and performing the Fourier transform $\hat{a}(\mathbf{r}) = (1/\sqrt{N}) \sum_\mathbf{k} \hat{a}_{\mathbf{k}} e^{i\mathbf{k}\cdot\mathbf{r}}$, where $\mathbf{r}$ is the position vector, $\mathbf{k}$ is the wave vector, and $N$ is the number of lattice sites, we obtain
\begin{align}
\mathcal{H}/\hbar =& \left[ \gamma H_0 \cos\theta + \gamma h_{\mathrm{d}} \cos(\omega_{\mathrm{d}} t) \sin\theta + \gamma A k^2 - i \alpha \tilde{\omega}_k \right] \hat{a}_\mathbf{k}^\dagger \hat{a}_\mathbf{k} \notag \\
& + \frac{\gamma h_0}{2} \left( e^{i\omega_0 t} + e^{-i\omega_0 t} \right) \left( \hat{a}_\mathbf{k}^\dagger + \hat{a}_\mathbf{k} \right), \label{eq.Hamil1}
\end{align}
where $k = |\mathbf{k}|$ is the wave vector magnitude, $\alpha$ is the damping coefficient, $h_0$ and $\omega_0$ are the amplitude and frequency of the injected signal, respectively, and $\tilde{\omega}_k$ is the renormalized magnon frequency defined below.

In the weak-field limit $h_{\mathrm{d}} \ll H_0$, we expand $\cos\theta$ and $\sin\theta$ up to second order in $h_{\mathrm{d}}$, yielding
\begin{equation}
\cos\theta \approx 1 - \frac{h_{\mathrm{d}}^2 \cos^2(\omega_{\mathrm{d}} t)}{2H_0^2}, \qquad
\sin\theta \approx \frac{h_{\mathrm{d}} \cos(\omega_{\mathrm{d}} t)}{H_0}.            \label{eq.triangle_appro}
\end{equation}
Substituting these expansions into Eq.~\eqref{eq.Hamil1}, we obtain
\begin{align}
\mathcal{H}/\hbar = \left( \tilde{\omega}_k + V\cos(2\omega_{\mathrm{d}} t) - i\alpha \tilde{\omega}_k \right) \hat{a}_\mathbf{k}^\dagger \hat{a}_\mathbf{k} + \Omega \left( e^{i\omega_0 t} + e^{-i\omega_0 t} \right) \left( \hat{a}_\mathbf{k}^\dagger + \hat{a}_\mathbf{k} \right), \label{eq.Hamil2}
\end{align}
where we have introduced the effective driving strength $\Omega = \gamma h_0/2$. The unperturbed magnon frequency is $\omega_k = \gamma H_0 + \gamma A k^2$. The time-independent part of the frequency acquires a static geometric shift $\Delta \omega_{\mathrm{dc}} = \gamma h_{\mathrm{d}}^2/(4H_0)$, yielding the renormalized frequency $\tilde{\omega}_k = \omega_k + \Delta \omega_{\mathrm{dc}}$. The time-dependent modulation is
\begin{equation}
V(t) = \frac{\gamma h_{\mathrm{d}}^2 \cos(2\omega_{\mathrm{d}} t)}{4H_0} \equiv V \cos(2\omega_{\mathrm{d}} t),
\end{equation}
with $V = \gamma h_{\mathrm{d}}^2/(4H_0)$. This modulation oscillates at $2\omega_{\mathrm{d}}$, explaining the comb spacing. For a fixed system length, the geometric factors are constants and are omitted hereafter.

To proceed, we apply the rotating-wave approximation (RWA) to Eq.~\eqref{eq.Hamil2}. We first perform the transformation
\begin{equation}
\hat{a}_\mathbf{k} \rightarrow \hat{a}_\mathbf{k} e^{-i\omega_0 t}, \qquad
\hat{a}_\mathbf{k}^\dagger \rightarrow \hat{a}_\mathbf{k}^\dagger e^{i\omega_0 t},
\end{equation}
and then neglect the rapidly oscillating terms proportional to $\hat{a}_\mathbf{k} e^{-i2\omega_0 t}$ and $\hat{a}_\mathbf{k}^\dagger e^{i2\omega_0 t}$. This yields

\begin{align}
    \mathcal{H} /\hbar =\left( \delta + V(t) -i\alpha \tilde{\omega}_k \right) \hat{a}_{\mathbf{k}}^{\dagger}\hat{a}_\mathbf{k}+\Omega\left( \hat{a}_{\mathbf{k}}^{\dagger}+\hat{a}_\mathbf{k} \right).               \label{Hamiltonian_final}
\end{align}

We define the effective detuning $\delta = \tilde{\omega}_k - \omega_0$. To determine the steady-state magnon population, we substitute the final Hamiltonian into the Heisenberg equation of motion, which gives
\begin{align}
    \frac{d\hat{a}_\mathbf{k}}{dt} = -i\left( \delta + V(t) - i\alpha \tilde{\omega}_k \right) \hat{a}_\mathbf{k} - i\Omega.
\end{align}
The solution to this differential equation can be expressed as
\begin{align}
    \Phi(t) &= -i\left( \delta - i\alpha \tilde{\omega}_k \right)t - i\beta \sin(2\omega_{\mathrm{d}} t), \\
    \hat{a}(t) &= \eta e^{\Phi(t)} \int_0^t e^{-\Phi(\tau)} d\tau,
\end{align}
where $\eta$ is a constant determined by the initial condition.

Employing the Jacobi-Anger expansion,
\begin{equation}
e^{-\Phi(t)} = e^{i(\delta - i\alpha \tilde{\omega}_k)t} \sum_{n=-\infty}^{\infty} J_n\!\left( \beta \right) e^{i2n\omega_{\mathrm{d}} t},
\end{equation}
we obtain the general solution
\begin{align}
\hat{a}(t) = -i\Omega & \left[ \sum_{m=-\infty}^{\infty} \sum_{n=-\infty}^{\infty} \frac{1}{i(2n\omega_{\mathrm{d}} + \delta - i\alpha \tilde{\omega}_k)} J_{-m}\!\left( \beta \right) J_n\!\left( \beta \right) e^{i2\omega_{\mathrm{d}}(n+m)t} \right. \notag \\
& \left. - \sum_{m=-\infty}^{\infty} \sum_{n=-\infty}^{\infty} \frac{e^{-i(\delta - i\alpha \tilde{\omega}_k)t}}{i(2n\omega_{\mathrm{d}} + \delta - i\alpha \tilde{\omega}_k)} J_{-m}\!\left( \beta \right) J_n\!\left( \beta \right) e^{i2\omega_{\mathrm{d}} m t} \right].
\end{align}
The second term contains a decaying factor $e^{-i(\delta - i\alpha \tilde{\omega}_k)t}$ (corresponding to the transient response) and is therefore neglected in the steady state. Defining $l = n + m$, the frequency of the $l$-th comb tooth is given by $\tilde{\omega}_k + 2l\omega_{\mathrm{d}}$, and its steady-state amplitude reads
\begin{equation}
\mathcal{A}_l = -\frac{\gamma h_0}{2} \sum_{n=-\infty}^{\infty} \frac{1}{\delta + 2n\omega_{\mathrm{d}} - i\alpha \tilde{\omega}_k} J_{n-l}\!\left( \beta \right) J_n\!\left( \beta \right).                        \label{eq.sideband_amp}
\end{equation}

\subsection{Theoretical framework II: Quantum treatment of the electromagnetic field}
\label{sec:quantum}

Building upon the previous derivation, where the Holstein-Primakoff transformation was performed in the local coordinate frame and the electromagnetic field was treated classically, we now develop an alternative quantum theory to describe the same physics. Following the same procedure as in the preceding section, we consider a spin chain in which spin waves are excited by a magnetic field of frequency $\omega_0$ in the excitation region. These spin waves subsequently propagate forward and enter the modulation region, where they are modulated by a linearly polarized magnetic field of frequency $\omega_{\mathrm{d}}$. Focusing on the modulation region and quantizing the electromagnetic field, we obtain the Hamiltonian
\begin{equation}
\mathcal{H}/\hbar = \omega_k \hat{a}_\mathbf{k}^\dagger \hat{a}_\mathbf{k} + \omega_{\mathrm{d}} \hat{c}^\dagger \hat{c} + \hat{H}_{\mathrm{int}},
\end{equation}
where $\omega_k$ and $\omega_{\mathrm{d}}$ are the frequencies of the unperturbed magnon and photon modes, respectively, $\hat{c}$ ($\hat{c}^\dagger$) is the photon annihilation (creation) operator, and $\hat{H}_{\mathrm{int}}$ accounts for their mutual interaction. In the linear regime, the magnon-photon coupling takes the standard form $\hat{H}_{\mathrm{int}} = g(\hat{c}^\dagger \hat{a}_\mathbf{k} + \hat{c} \hat{a}_\mathbf{k}^\dagger)$, where $g$ is the linear magnon-photon coupling strength. However, by keeping higher-order terms in the Holstein-Primakoff transformation~\cite{HolsteinPrimakoff1940,Dyson1956General}, additional nonlinear coupling channels arise. In the laboratory frame, expanding the Holstein-Primakoff transformation to second order gives
\begin{align}
\hat{S}^+ &= \sqrt{2S} \left( 1 - \frac{1}{4S} \hat{a}^\dagger \hat{a} \right) \hat{a}, \notag \\
\hat{S}^- &= \sqrt{2S} \hat{a}^\dagger \left( 1 - \frac{1}{4S} \hat{a}^\dagger \hat{a} \right), \notag \\
\hat{S}_z &= S - \hat{a}^\dagger \hat{a}, \label{eq.nonlinear_HP}
\end{align}
from which the Cartesian spin components follow as
\begin{align}
\hat{S}_x &= \frac{1}{2} \left( \hat{S}^+ + \hat{S}^- \right), \notag \\
\hat{S}_y &= \frac{1}{2i} \left( \hat{S}^+ - \hat{S}^- \right),
\end{align}
where $S$ is the spin quantum number.

For a linearly polarized modulation field $\mathbf{B}_{\mathrm{d}}(t) = \mu_0 h_{\mathrm{d}} \cos(\omega_{\mathrm{d}} t) \hat{\mathbf{x}}$, the quantized field takes the form
\begin{equation}
\hat{\mathbf{B}}_{\mathrm{d}}(t) = \mu_0 h_{\mathrm{d}} \left( \hat{\boldsymbol{\epsilon}}_+ \hat{c}_+ e^{-i\omega_{\mathrm{d}} t} + \hat{\boldsymbol{\epsilon}}_- \hat{c}_- e^{-i\omega_{\mathrm{d}} t} + \mathrm{h.c.} \right), \label{eq.linear_photon}
\end{equation}
where $\hat{\boldsymbol{\epsilon}}_\pm = (\hat{\mathbf{x}} \pm i\hat{\mathbf{y}})/\sqrt{2}$ are the circular polarization basis vectors, and $\hat{c}_\pm$ ($\hat{c}_\pm^\dagger$) are the photon annihilation (creation) operators for the two circular polarizations. Since the field is linearly polarized along $\hat{\mathbf{x}}$, we are primarily concerned with the $B_x$ component and its coupling to the magnetization $m_x$. Extracting the $\hat{\mathbf{x}}$ component from Eq.~\eqref{eq.linear_photon} gives
\begin{equation}
B_x(t) = \mu_0 h_{\mathrm{d}} \left( \hat{c}_+ e^{-i\omega_{\mathrm{d}} t} + \hat{c}_+^\dagger e^{i\omega_{\mathrm{d}} t} + \hat{c}_- e^{-i\omega_{\mathrm{d}} t} + \hat{c}_-^\dagger e^{i\omega_{\mathrm{d}} t} \right). \label{eq.linear_photon_Bx}
\end{equation}
This expression can be simplified by introducing the linear-polarization operators $\hat{c}_l = \frac{\hat{c}_+ + \hat{c}_-}{\sqrt{2}}$ and $\hat{c}_l^\dagger = \hat{c}_+^\dagger + \hat{c}_-^\dagger$. Substituting these definitions into Eq.~\eqref{eq.linear_photon_Bx}, we arrive at the compact form
\begin{equation}
\mathbf{B}_{\mathrm{d}}(t) = \mu_0 h_{\mathrm{d}} \left( \hat{c}_l e^{-i\omega_{\mathrm{d}} t} + \hat{c}_l^\dagger e^{i\omega_{\mathrm{d}} t} \right).
\end{equation}

We proceed to evaluate the magnon-photon interaction Hamiltonian. To second order in the Holstein-Primakoff expansion, the interaction terms read
\begin{align}
\hat{H}_{\mathrm{int}}^{(1)} &= \frac{\gamma\sqrt{2S}\mu_0 h_{\mathrm{d}}}{2} \int \left( \hat{a} \hat{c}_l^\dagger e^{i\omega_{\mathrm{d}} t} + \hat{a} \hat{c}_l e^{-i\omega_{\mathrm{d}} t} + \mathrm{h.c.} \right) d^3r, \label{cm_Hamiltionian_1} \\
\hat{H}_{\mathrm{int}}^{(2)} &= -\frac{\gamma\sqrt{2S}\mu_0 h_{\mathrm{d}}}{8S} \int \left( \hat{c}_l^\dagger e^{i\omega_{\mathrm{d}} t} \hat{a}^\dagger \hat{a} \hat{a} + \hat{c}_l e^{-i\omega_{\mathrm{d}} t} \hat{a}^\dagger \hat{a} \hat{a} + \mathrm{h.c.} \right) d^3r. \label{cm_Hamiltionian_2}
\end{align}
Note that the Hermitian conjugate of the operator products must be handled carefully: since $(ABC)^\dagger = C^\dagger B^\dagger A^\dagger$, we have $(\hat{a}^\dagger \hat{a} \hat{a})^\dagger = \hat{a}^\dagger \hat{a}^\dagger \hat{a}$. In conventional treatments, the term $\hat{a} \hat{c}_l e^{-i\omega_{\mathrm{d}} t}$ is often neglected as a fast-oscillating component. Nevertheless, because $\omega_{\mathrm{d}} \ll \omega_k$ in our setup, such a rotating-wave approximation is not justified, and we therefore keep this term. Performing the Fourier expansion $\hat{a}(\mathbf{r}) = (1/\sqrt{N}) \sum_\mathbf{k} \hat{a}_{\mathbf{k}} e^{i\mathbf{k}\cdot\mathbf{r}}$, we obtain the corresponding momentum-space Hamiltonian.

\begin{align}
\hat{H}_{\mathrm{int}}^{(1)} &= -\frac{\gamma\sqrt{2S}\mu_0 h_{\mathrm{d}}}{2N^{1/2}} \sum_{\mathbf{k}_1} \int \left[ \left( \hat{a}_{\mathbf{k}_1} \hat{c}_l^\dagger e^{i\omega_{\mathrm{d}} t} + \hat{a}_{\mathbf{k}_1} \hat{c}_l e^{-i\omega_{\mathrm{d}} t} \right) e^{i\mathbf{k}_1 \cdot \mathbf{r}} + \mathrm{h.c.} \right] d^3r, \label{Fourier_cm_Hamiltionian_1} \\
\hat{H}_{\mathrm{int}}^{(2)} &= \frac{\gamma\sqrt{2S}\mu_0 h_{\mathrm{d}}}{8S N^{3/2}} \sum_{\mathbf{k}_1, \mathbf{k}_2, \mathbf{k}_3} \int \left[ \left( \hat{c}_l^\dagger e^{i\omega_{\mathrm{d}} t} \hat{a}_{\mathbf{k}_3}^\dagger \hat{a}_{\mathbf{k}_2} \hat{a}_{\mathbf{k}_1} + \hat{c}_l e^{-i\omega_{\mathrm{d}} t} \hat{a}_{\mathbf{k}_3}^\dagger \hat{a}_{\mathbf{k}_2} \hat{a}_{\mathbf{k}_1} \right) e^{i(\mathbf{k}_1 + \mathbf{k}_2 - \mathbf{k}_3) \cdot \mathbf{r}} + \mathrm{h.c.} \right] d^3r. \label{Fourier_cm_Hamiltionian_2}
\end{align}

Since $\omega_{\mathrm{d}} \ll \gamma H_0$, the driving field cannot efficiently excite spin-wave modes directly. From the classical model, we know that the background magnetization $\mathbf{m}_0$ undergoes a rigid motion driven by the low-frequency modulation field, and this motion is uniform within the modulation region. We therefore set $\mathbf{k}_1 = 0$, and the corresponding operator $\hat{a}_{\mathbf{k}_1}$ describes the collective motion mode of the background magnetization, i.e., the coherent $\mathbf{k}=0$ mode. From Eq.~\eqref{Fourier_cm_Hamiltionian_1}, we infer that $\hat{a}_0(t) = \hat{a}^{(+)} e^{-i\omega_{\mathrm{d}} t} + \hat{a}^{(-)} e^{i\omega_{\mathrm{d}} t}$. For a linearly polarized driving field, the amplitudes satisfy $\langle \hat{a}^{(+)} \rangle = \langle \hat{a}^{(-)} \rangle$. Substituting $\hat{a}_0(t) = \hat{a}^{(+)} e^{-i\omega_{\mathrm{d}} t} + \hat{a}^{(-)} e^{i\omega_{\mathrm{d}} t}$ for $\hat{a}_{\mathbf{k}_1}$ in Eq.~\eqref{Fourier_cm_Hamiltionian_2}, we obtain
\begin{align}
\hat{H}_{\mathrm{int}}^{(2)} 
&= \frac{\gamma\sqrt{2S}\mu_0 h_{\mathrm{d}}}{8S N^{3/2}} \notag \\
&\quad \times \sum_{\mathbf{k}_2, \mathbf{k}_3} \int \Big[ \left( \hat{c}_l^\dagger \hat{a}^{(+)} + \hat{c}_l^\dagger \hat{a}^{(-)} e^{i2\omega_{\mathrm{d}} t} + \hat{c}_l \hat{a}^{(+)} e^{-i2\omega_{\mathrm{d}} t} + \hat{c}_l \hat{a}^{(-)} \right) \hat{a}_{\mathbf{k}_3}^\dagger \hat{a}_{\mathbf{k}_2} e^{i(\mathbf{k}_2 - \mathbf{k}_3) \cdot \mathbf{r}} + \mathrm{h.c.} \Big] d^3r.
\end{align}

At this stage, we observe that $\hat{H}_{\mathrm{int}}^{(2)}$ is periodic with period $T = \pi/\omega_{\mathrm{d}}$. Since both $\hat{a}^{(\pm)}$ and $\hat{c}_l$ are spatially uniform within the modulation region, the integral is nonzero only when $\mathbf{k}_2 = \mathbf{k}_3$. To simplify the second-order interaction Hamiltonian, we introduce the operators
\begin{equation}
\hat{B} = \hat{c}_l^\dagger \hat{a}^{(+)} + \hat{c}_l \hat{a}^{(-)}, \qquad
\hat{A} = \hat{c}_l^\dagger \hat{a}^{(-)} e^{i2\omega_{\mathrm{d}} t} + \hat{c}_l \hat{a}^{(+)} e^{-i2\omega_{\mathrm{d}} t},
\end{equation}
and define $g_2 = -\frac{\gamma\sqrt{2S}\mu_0 h_{\mathrm{d}}}{8S N^{3/2}}$. The Hamiltonian then becomes
\begin{equation}
\hat{H}_{\mathrm{int}}^{(2)} = g_2 \sum_{\mathbf{k}_2} \left( \hat{A} \hat{a}_{\mathbf{k}_2}^\dagger \hat{a}_{\mathbf{k}_2} + \hat{B} \hat{a}_{\mathbf{k}_2}^\dagger \hat{a}_{\mathbf{k}_2} + \mathrm{h.c.} \right).
\end{equation}
Since $\hat{a}_{\mathbf{k}_2}^\dagger \hat{a}_{\mathbf{k}_2}$ is Hermitian and the expectation values of $\hat{A}$ and $\hat{B}$ are real in the steady state, the Hermitian conjugate (h.c.) simply doubles the coefficients when taking the mean-field expectation value. Applying the mean-field approximation to replace $\hat{A}$ and $\hat{B}$ by their expectation values
\begin{equation}
\langle \hat{B} \rangle = \langle \hat{c}_l^\dagger\rangle  \langle \hat{a}^{(+)} \rangle + \langle \hat{c}_l\rangle  \langle \hat{a}^{(-)}\rangle , \qquad
\langle \hat{A} \rangle = \langle \hat{c}_l^\dagger\rangle  \langle \hat{a}^{(-)}\rangle  e^{i2\omega_{\mathrm{d}} t} + \langle \hat{c}_l\rangle  \langle \hat{a}^{(+)}\rangle  e^{-i2\omega_{\mathrm{d}} t},
\end{equation}
and omitting the summation over $\mathbf{k}_2$, we finally express the second-order interaction Hamiltonian in the compact form
\begin{equation}
\hat{H}_{\mathrm{int}}^{(2)} = G \hat{a}_\mathbf{k}^\dagger \hat{a}_\mathbf{k} + \frac{V}{2} e^{-i2\omega_{\mathrm{d}} t} \hat{a}_\mathbf{k}^\dagger \hat{a}_\mathbf{k} + \frac{V}{2} e^{i2\omega_{\mathrm{d}} t} \hat{a}_\mathbf{k}^\dagger \hat{a}_\mathbf{k}.
\end{equation}
This effective interaction is diagonal in momentum space and proportional to $\hat{a}^\dagger\hat{a}$, representing a time-dependent two-magnon parametric modulation $V\cos(2\omega_{\mathrm{d}} t)\hat{a}^\dagger\hat{a}$. The explicit decoupling between the external drive amplitude and the injected signal confirms that the seed spin wave can remain in the linear regime while the comb is generated.

Here, the effective coupling constants are given by
\begin{equation}
G = V = -\frac{\gamma\sqrt{2S}\mu_0 h_{\mathrm{d}}}{4S N^{3/2}} \langle \hat{c}_l \rangle \langle \hat{a}^{(\pm)} \rangle,
\end{equation}
which are consistent with the results obtained from the classical model. This agreement indicates that one could directly solve the differential equations following the classical approach. Nevertheless, we adopt the Floquet formalism here to provide a more systematic treatment. Considering the full system Hamiltonian, with a spin-wave input at frequency $\omega_0$ and a modulation field at $\omega_{\mathrm{d}}$, and omitting the photon energy contribution, we have
\begin{equation}
\mathcal{H}/\hbar = (\omega_k + G) \hat{a}_\mathbf{k}^\dagger \hat{a}_\mathbf{k} + \frac{V}{2} e^{-i2\omega_{\mathrm{d}} t} \hat{a}_\mathbf{k}^\dagger \hat{a}_\mathbf{k} + \frac{V}{2} e^{i2\omega_{\mathrm{d}} t} \hat{a}_\mathbf{k}^\dagger \hat{a}_\mathbf{k} + \Omega \left( \hat{a}_\mathbf{k} e^{i\omega_0 t} + \hat{a}_\mathbf{k}^\dagger e^{-i\omega_0 t} \right),
\end{equation}
where $\Omega = \gamma h_0/2$ is the driving strength defined previously. The renormalized magnon frequency is $\tilde{\omega}_k = \omega_k + G$, where $G$ is the static geometric shift derived above (corresponding to $\Delta \omega_{\mathrm{dc}}$ in the classical picture). We define the effective detuning $\delta = \tilde{\omega}_k - \omega_0$. After applying the rotating-wave transformation, the Hamiltonian becomes
\begin{equation}
\mathcal{H}/\hbar = \delta \hat{a}_\mathbf{k}^\dagger \hat{a}_\mathbf{k} + \frac{V}{2} e^{-i2\omega_{\mathrm{d}} t} \hat{a}_\mathbf{k}^\dagger \hat{a}_\mathbf{k} + \frac{V}{2} e^{i2\omega_{\mathrm{d}} t} \hat{a}_\mathbf{k}^\dagger \hat{a}_\mathbf{k} + \Omega \left( \hat{a}_\mathbf{k} + \hat{a}_\mathbf{k}^\dagger \right). \label{Floquet_Hamiltionian}
\end{equation}
Substituting the Hamiltonian in Eq.~\eqref{Floquet_Hamiltionian} into the Heisenberg equation of motion yields
\begin{align}
    \frac{d\hat{a}_\mathbf{k}}{dt} = -i\left[ \delta \hat{a}_\mathbf{k} + \frac{V}{2} e^{-i2\omega_{\mathrm{d}} t} \hat{a}_\mathbf{k} + \frac{V}{2} e^{i2\omega_{\mathrm{d}} t} \hat{a}_\mathbf{k} + \Omega \right]. \label{Heisenberg}
\end{align}

In the absence of driving, one may assume $\hat{a}(t) = e^{-i\varepsilon t} u(t)$, where $\varepsilon$ is the Floquet quasienergy and $u(t) = u(t+T)$ shares the same periodicity as the Hamiltonian\cite{shirley1965solution}. In the present problem, however, the driving term $\Omega$ is present. If $\varepsilon$ is chosen not to be an integer multiple of $2\omega_{\mathrm{d}}$, the driving term acquires an additional factor $e^{i\varepsilon t}$, which is not periodic with $T = \pi/\omega_{\mathrm{d}}$. This breaks the freedom to arbitrarily choose the Floquet phase $e^{-i\varepsilon t}$. Without loss of generality, we set $\varepsilon = 0$. Following the Floquet formalism, the solution to the Heisenberg equation of motion (Eq.~\ref{Heisenberg}) takes the form $\hat{a}(t) = \sum_n a^{(n)} e^{-i n 2\omega_{\mathrm{d}} t}$, and the Hamiltonian is expanded as $H(t) = \sum_m H^{(m)} e^{-i m 2\omega_{\mathrm{d}} t}$. Substituting these two periodic expansions into Eq.~\ref{Heisenberg} and comparing the coefficients of each Fourier component, we obtain the equation satisfied by the $n$-th component $a^{(n)}$:
\begin{equation}
2n\omega_{\mathrm{d}} a^{(n)} = \sum_m H^{(m)} a^{(n-m)} + \Omega \delta_{n0}.
\end{equation}
Rearranging, we define the Floquet Hamiltonian with matrix elements
\begin{equation}
H_{\mathrm{F}(nm)} = n 2\omega_{\mathrm{d}} \delta_{nm} - H^{(n-m)},
\end{equation}
so that the Floquet equation reads $H_{\mathrm{F}} \mathbf{a} = \mathbf{C}$, where $\mathbf{a} = (\cdots, a^{(-1)}, a^{(0)}, a^{(1)}, \cdots)^T$ and $\mathbf{C} = (\cdots, 0, \Omega, 0, \cdots)^T$. The above discussion neglects energy dissipation in the system. In this work, since no probing process is involved, we only consider the Gilbert damping, characterized by the damping factor $\alpha$.

When Gilbert damping is included, the Floquet equation takes the form
\begin{align}
\left( H_{\mathrm{F}} + i\alpha \tilde{\omega}_k \mathbf{I} \right) \mathbf{a} = \mathbf{C},
\end{align}
where $\mathbf{I}$ is the identity matrix. Since the Hamiltonian contains only the Fourier components $H^{(0)}$ and $H^{(\pm 1)}$, the resulting Floquet problem maps exactly onto a tight-binding Wannier-Stark ladder. Here $H_{\mathrm{F}}$ is the undamped Floquet Hamiltonian, and the Gilbert damping is included through the term $i\alpha\tilde{\omega}_k\mathbf{I}$. The Green's function $G^{\mathrm{F}} = (H_{\mathrm{F}} + i\alpha \tilde{\omega}_k \mathbf{I})^{-1}$ admits an analytical solution, with its matrix element $G^{\mathrm{F}}_{mn}$ describing the steady-state response of the $n$-th Floquet replica to a unit drive applied at the $m$-th sideband. Evaluating the response function $G^{\mathrm{F}}_{l0}$ at the $l$-th comb tooth frequency $\omega_l = \tilde{\omega}_k + 2l\omega_{\mathrm{d}}$ yields
\begin{equation}
G^{\mathrm{F}}_{l0} = \sum_{n=-\infty}^{\infty} \frac{1}{\delta + 2n\omega_{\mathrm{d}} - i\alpha \tilde{\omega}_k} J_{n-l}\!\left( \beta \right) J_n\!\left( \beta \right).
\end{equation}
The physical amplitude of the $l$-th sideband, incorporating the driving term from $\mathbf{C}$, is given by $\mathcal{A}_l = -(\gamma h_0/2) G^{\mathrm{F}}_{l0}$, which evaluates to
\begin{equation}
\mathcal{A}_l = -\frac{\gamma h_0}{2} \sum_{n=-\infty}^{\infty} \frac{1}{\delta + 2n\omega_{\mathrm{d}} - i\alpha \tilde{\omega}_k} J_{n-l}\!\left( \beta \right) J_n\!\left( \beta \right).
\end{equation}
This expression for the sideband amplitude reproduces the classical result exactly.

\subsection{Discussion: higher-order geometric harmonics and spectral flatness}
An important advantage of the transverse driving scheme over the conventional longitudinal configuration lies in the inherent anharmonicity of the geometric constraint~\cite{Dyson1956General,Dyson1956Thermodynamic}. To quantify this, we expand the effective longitudinal modulation beyond the leading-order term. The exact instantaneous frequency is $\omega_k^\perp(t) = \gamma \sqrt{H_0^2 + h_{\mathrm{d}}^2 \cos^2(\omega_{\mathrm{d}} t)}$. Expanding to $O(h_{\mathrm{d}}^4)$ yields
\begin{equation}
\omega_k^\perp(t) = \tilde{\omega}_k + V \cos(2\omega_{\mathrm{d}} t) + V_4 \cos(4\omega_{\mathrm{d}} t) + V_6 \cos(6\omega_{\mathrm{d}} t) + \cdots,
\end{equation}
with $V = \gamma h_{\mathrm{d}}^2/(4H_0)$, $V_4 = -\gamma h_{\mathrm{d}}^4/(64 H_0^3)$, and $V_6 = \gamma h_{\mathrm{d}}^6/(512 H_0^5)$. These multi-frequency modulations produce additional sidebands at multiples of $4\omega_{\mathrm{d}}$ and $6\omega_{\mathrm{d}}$, effectively filling the deep dips of the single-frequency Bessel envelope and leading to a flatter overall spectrum at moderate to strong driving fields.

\subsection{Superposition of driving fields with different polarizations}
In the preceding section, we considered the case of a linearly polarized driving field. We now turn to more general driving configurations, specifically: (1) a superposition of two circularly polarized fields with the same helicity, and (2) a superposition of two circularly polarized fields with opposite helicities. The resulting frequency comb spacings are $\omega_{\mathrm{d}1} - \omega_{\mathrm{d}2}$ and $\omega_{\mathrm{d}1} + \omega_{\mathrm{d}2}$, respectively.

To understand this from the perspective of the classical electromagnetic Hamiltonian, we adopt the adiabatic approximation and neglect the motion of the macroscopic magnetization $\mathbf{M}_0$, considering only the time dependence of its magnitude. The total effective field can then be written as
\begin{equation}
H_{\mathrm{tot}} = \sqrt{ H_0^2 + \left( h_{\mathrm{d}1} \cos \omega_{\mathrm{d}1} t + h_{\mathrm{d}2} \cos \omega_{\mathrm{d}2} t \right)^2 + \left( h_{\mathrm{d}1} \sin \omega_{\mathrm{d}1} t + \sigma h_{\mathrm{d}2} \sin \omega_{\mathrm{d}2} t \right)^2 },
\end{equation}
where $\sigma = +1$ corresponds to the case where the two driving fields have the same helicity (co-rotating), and $\sigma = -1$ corresponds to the case where they have opposite helicities (counter-rotating). Using trigonometric identities, the total effective field oscillates at the frequency $\omega_{\mathrm{d}1} - \sigma \omega_{\mathrm{d}2}$. According to Floquet theory, the solution then contains frequency components at integer multiples of $\omega_{\mathrm{d}1} - \sigma \omega_{\mathrm{d}2}$, which precisely explains the observed comb spacing.

For a quantum treatment, we need the explicit form of the photon operators. To avoid the ambiguity associated with the terms ``left-handed'' and ``right-handed,'' we label the photon polarization states by their helicity $\sigma = \pm 1$, defined as the projection of the photon spin angular momentum onto the propagation direction ($+\hat{\mathbf{z}}$), in units of $\hbar$. Correspondingly, the circular polarization basis vectors are chosen as
\[
\hat{\boldsymbol{\epsilon}}_{\sigma} = \hat{\mathbf{x}} + i\sigma \hat{\mathbf{y}}, \qquad \sigma = \pm 1,
\]
where $\sigma = +1$ denotes helicity $+1$ (electric field rotating counterclockwise when viewed along $+\hat{\mathbf{z}}$), and $\sigma = -1$ denotes helicity $-1$ (clockwise rotation).

However, under the Holstein-Primakoff transformation, we have
\begin{align}
\langle \hat{S}_x \rangle &= \frac{\sqrt{2S}}{2} a_0 \left( e^{-i\omega t} + e^{i\omega t} \right) = \sqrt{2S} a_0 \cos \omega t, \notag \\
\langle \hat{S}_y \rangle &= \frac{\sqrt{2S}}{2i} a_0 \left( e^{-i\omega t} - e^{i\omega t} \right) = -\sqrt{2S} a_0 \sin \omega t.
\end{align}
This indicates that the actual rotation of the angular momentum $\hat{\mathbf{S}}$ follows $\cos{\omega t}\,\hat{\mathbf{x}} - \sin{\omega t}\,\hat{\mathbf{y}}$, which is opposite to the LLG convention. This reversal arises because, in the Holstein-Primakoff representation, the spin raising operator $\hat{S}_+$ is identified with the magnon annihilation operator $\hat{a}$, effectively taking the complex conjugate and thus reversing the sense of magnetization precession in the quantum picture relative to the classical one.

The quantized electromagnetic field can therefore be written in a unified form as
\begin{align}
\hat{\mathbf{B}}_+ &= \hat{\boldsymbol{\epsilon}}_{+} \hat{c}_{+} e^{-i\omega_{\mathrm{d}1} t} + \hat{\boldsymbol{\epsilon}}_{-} \hat{c}_{+}^{\dagger} e^{i\omega_{\mathrm{d}1} t}, \\
\hat{\mathbf{B}}_- &= \hat{\boldsymbol{\epsilon}}_{-} \hat{c}_{-} e^{-i\omega_{\mathrm{d}2} t} + \hat{\boldsymbol{\epsilon}}_{+} \hat{c}_{-}^{\dagger} e^{i\omega_{\mathrm{d}2} t},
\end{align}
where $\hat{c}_{\sigma}$ ($\hat{c}_{\sigma}^{\dagger}$) is the annihilation (creation) operator for a photon with helicity $\sigma$. The $\hat{\mathbf{x}}$ and $\hat{\mathbf{y}}$ components of the magnetic field are given by
\begin{align}
\hat{B}_x &= \hat{c}_+ e^{-i\omega_{\mathrm{d}1} t} + \hat{c}_{+}^{\dagger} e^{i\omega_{\mathrm{d}1} t} + \hat{c}_{-} e^{-i\omega_{\mathrm{d}2} t} + \hat{c}_-^{\dagger} e^{i\omega_{\mathrm{d}2} t}, \\
\hat{B}_y &= i\left( \hat{c}_+ e^{-i\omega_{\mathrm{d}1} t} - \hat{c}_{+}^{\dagger} e^{i\omega_{\mathrm{d}1} t} - \hat{c}_{-} e^{-i\omega_{\mathrm{d}2} t} + \hat{c}_-^{\dagger} e^{i\omega_{\mathrm{d}2} t} \right).
\end{align}

Still working within the laboratory-frame Holstein-Primakoff transformation given in Eq.~\eqref{eq.nonlinear_HP}, the first-order interaction Hamiltonian is obtained as $H_{\mathrm{int}}^{(1)} = -\gamma(\hat{S}_{x}^{(1)} \hat{B}_x + \hat{S}_{y}^{(1)} \hat{B}_y)$, which explicitly reads
\begin{align}
\hat{H}_{\mathrm{int}}^{(1)}
=
-\gamma\sqrt{2S}\big(
&\hat{a}\hat{c}_{+}e^{-i\omega_{d1}t}
+\hat{a}\hat{c}_{-}^{\dagger}e^{i\omega_{d2}t}
+\hat{a}^{\dagger}\hat{c}_{+}^{\dagger}e^{i\omega_{d1}t}
+\hat{a}^{\dagger}\hat{c}_{-}e^{-i\omega_{d2}t}
\big).
\end{align}
In conventional treatments, only certain terms are considered physical, reflecting the well-known selection rule that magnon excitations couple selectively to photon helicity. However, in our case, both $\omega_{\mathrm{d}1}$ and $\omega_{\mathrm{d}2}$ are much lower than the typical magnon precession frequencies (in the GHz range), and the excited modes correspond to collective deformations of the background magnetization rather than propagating spin waves. Consequently, the counter-rotating terms also acquire physical significance, as they describe the excitation of collective modes. The operator for the collective background magnetization mode can therefore be written, in the same convention as in the main text (annihilation operators associated with $e^{-i\omega t}$ and creation operators with $e^{+i\omega t}$), as follows:
\begin{equation}
\hat{a}_0 = \hat{a}_{0}^{(1)} e^{i\omega_{\mathrm{d}1} t} + \hat{a}_{0}^{(2)} e^{-i\omega_{\mathrm{d}2} t},
\end{equation}
together with its Hermitian conjugate
\begin{equation}
\hat{a}_0^\dagger = \hat{a}_{0}^{(1)\dagger} e^{-i\omega_{\mathrm{d}1} t} + \hat{a}_{0}^{(2)\dagger} e^{i\omega_{\mathrm{d}2} t}.
\end{equation}

Proceeding to the second-order interaction, we have
\begin{align}
\hat{H}_{\mathrm{int}}^{(2)}
={}&
\frac{\gamma\sqrt{2S}}{4S}
\hat{a}^{\dagger}\hat{a}
\Big[
\hat{a}_{0}
\left(
\hat{c}_{+}e^{-i\omega_{\mathrm{d}1}t}
+
\hat{c}_{-}^{\dagger}e^{i\omega_{\mathrm{d}2}t}
\right)
+
\hat{a}_{0}^{\dagger}
\left(
\hat{c}_{+}^{\dagger}e^{i\omega_{\mathrm{d}1}t}
+
\hat{c}_{-}e^{-i\omega_{\mathrm{d}2}t}
\right)
\Big]
\nonumber\\
={}&
\frac{\gamma\sqrt{2S}}{4S}
\hat{a}^{\dagger}\hat{a}
\Big\{
\hat{a}_{0}^{(1)}\hat{c}_{+}
+
\hat{a}_{0}^{(2)}\hat{c}_{-}^{\dagger}
+
\hat{a}_{0}^{(1)\dagger}\hat{c}_{+}^{\dagger}
+
\hat{a}_{0}^{(2)\dagger}\hat{c}_{-}
\nonumber\\
&\qquad
+
\left(
\hat{a}_{0}^{(1)}\hat{c}_{-}^{\dagger}
+
\hat{a}_{0}^{(2)\dagger}\hat{c}_{+}^{\dagger}
\right)
e^{i(\omega_{\mathrm{d}1}+\omega_{\mathrm{d}2})t}
\nonumber\\
&\qquad
+
\left(
\hat{a}_{0}^{(1)\dagger}\hat{c}_{-}
+
\hat{a}_{0}^{(2)}\hat{c}_{+}
\right)
e^{-i(\omega_{\mathrm{d}1}+\omega_{\mathrm{d}2})t}
\Big\}.
\end{align}

Analogously, by introducing
\begin{equation}
\hat{A} = \hat{a}_{0}^{(2)} \hat{c}_+ e^{-i(\omega_{\mathrm{d}2}+\omega_{\mathrm{d}1}) t} + \hat{a}_{0}^{(1)} \hat{c}^{\dagger}_- e^{i(\omega_{\mathrm{d}2}+\omega_{\mathrm{d}1}) t}, \qquad
\hat{B} = \hat{a}_{0}^{(1)} \hat{c}_+ + \hat{a}_{0}^{(2)} \hat{c}^{\dagger}_-,
\end{equation}
the second-order Hamiltonian is found to be periodic with period $T = 2\pi/(\omega_{\mathrm{d}1}+\omega_{\mathrm{d}2})$, indicating that the resulting frequency comb spacing is $\omega_{\mathrm{d}1}+\omega_{\mathrm{d}2}$. The same-helicity case is completely analogous, with the sum frequency replaced by $|\omega_{\mathrm{d}1}-\omega_{\mathrm{d}2}|$, corresponding to the difference-frequency comb spacing discussed in the main text. Within the mean-field approximation, the sideband intensities can then be computed using the Floquet formalism.

\section{FLOQUET STATES IN THE DISPERSION RELATION}
\label{sec:floquet_dispersion}

\begin{figure*}[t]
    \centering
    \includegraphics[width=0.9\textwidth]{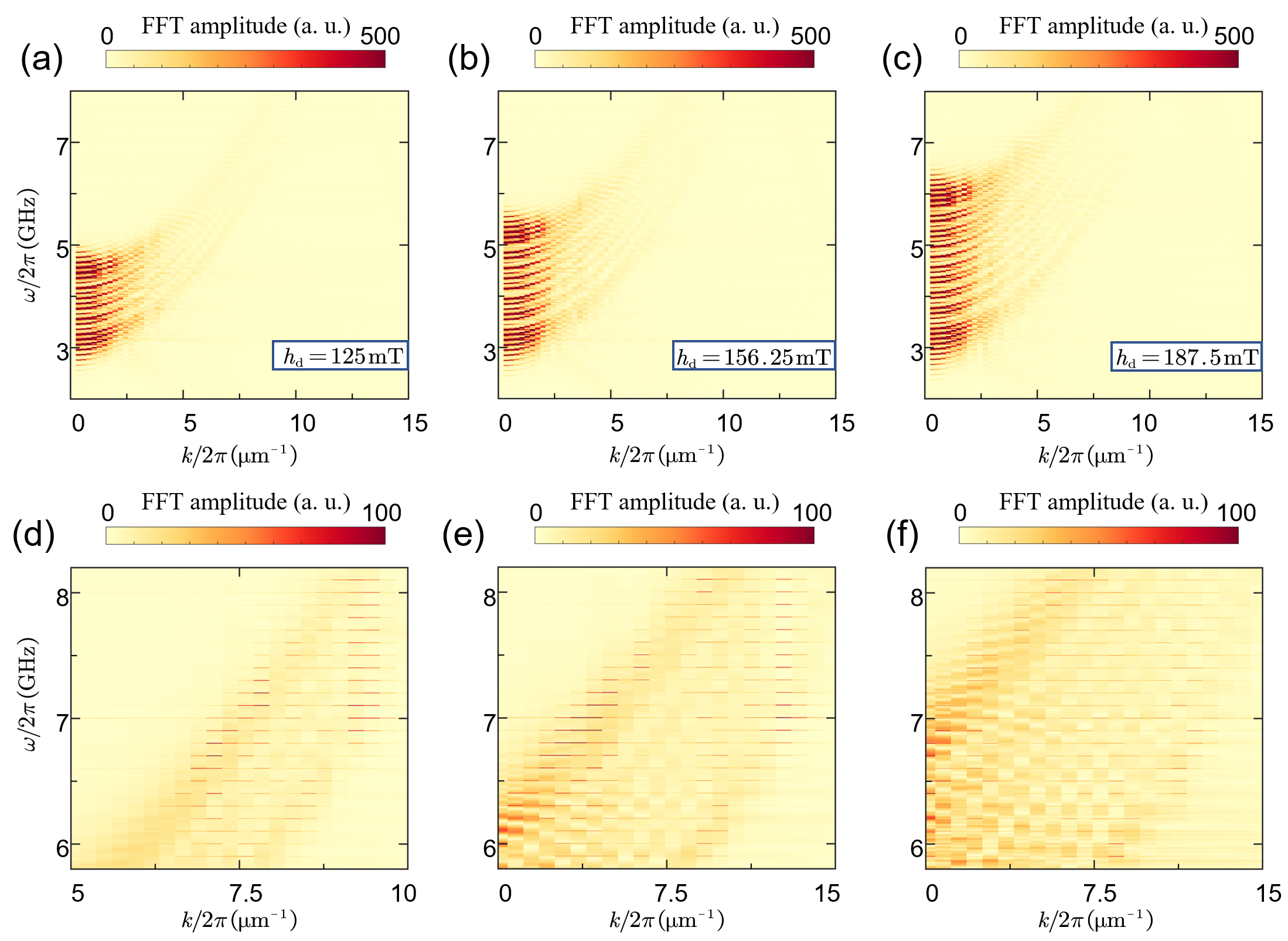}
    \caption{Floquet states and momentum shift. (a)-(c) Evolution of Floquet sidebands and the upward shift of dispersion branches with increasing driving amplitude. (d)-(f) Magnified views at the injected spin-wave frequency of 7 GHz, showing the emergence of the MFC and the corresponding decrease in momentum.}
    \label{fig:SM_fig2}
\end{figure*}

To better understand the Floquet properties of the present system, we extended the modulation region in our COMSOL simulations and extracted the dispersion relation within this region. The results are presented in Fig.~\ref{fig:SM_fig2}. Panels (a)-(c) show the evolution of the Floquet sidebands as the driving amplitude increases, where higher-order sidebands progressively emerge. In addition, the entire dispersion branches are shifted upward, reflecting the increase in the ground-state energy induced by the periodic driving. Panels (d)-(f) provide a magnified view of the 7 GHz region indicated in panels (a)-(c). At this frequency, the spin-wave excitations clearly give rise to a frequency comb. Moreover, with the upward shift of the dispersion curves and the proliferation of sidebands, the momentum associated with the comb teeth is found to decrease systematically. This momentum shift refers to the bare dispersion relation; the Floquet quasi-momentum index itself is unchanged by the diagonal modulation.

\section{MUMAX3 SIMULATIONS}
\label{sec:mumax}

\begin{figure*}[t]
    \centering
    \includegraphics[width=0.9\textwidth]{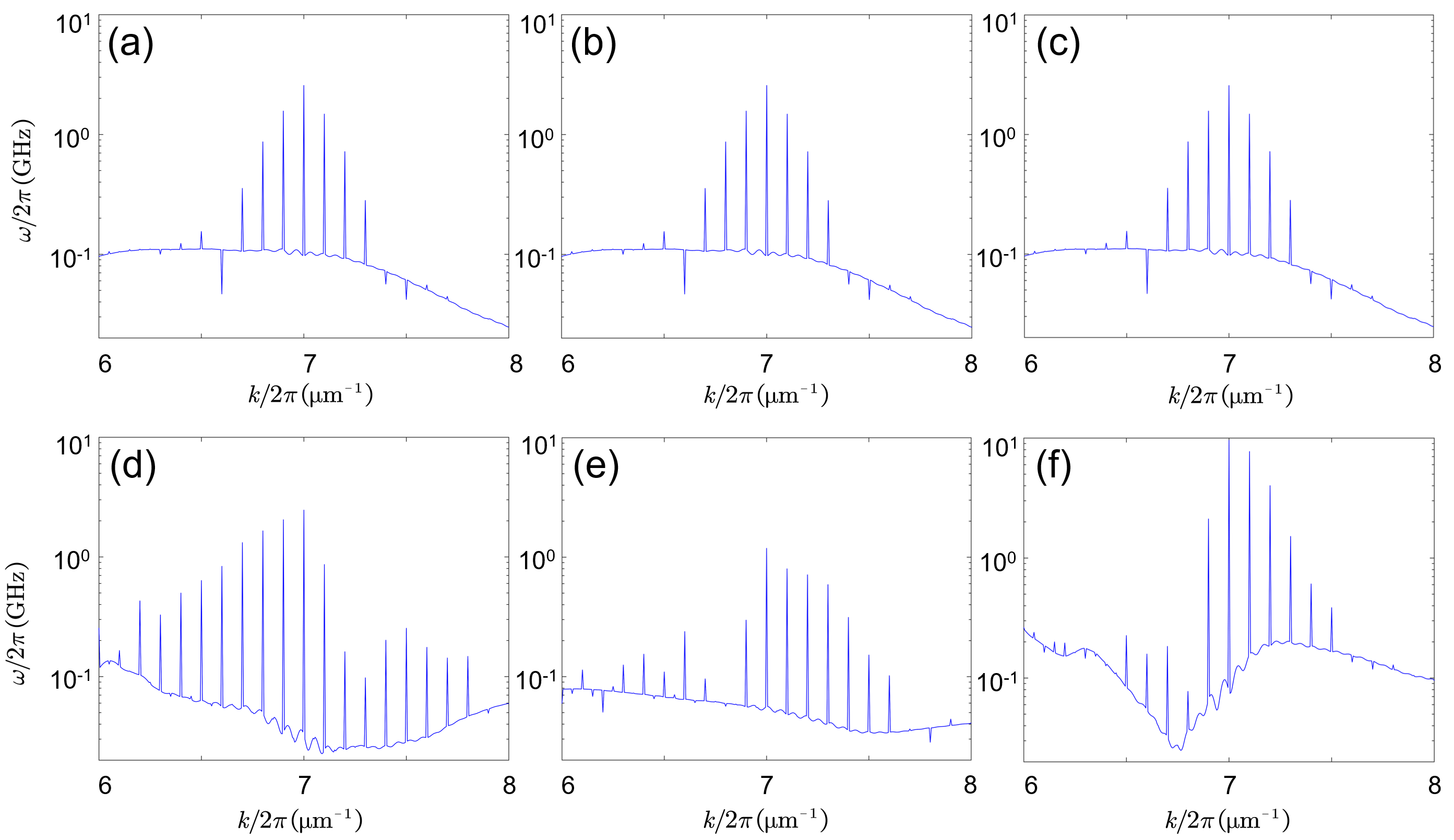}
    \caption{Mumax3 simulations of the MFC in a thin-film system. (a)-(c) Results for backward-volume, forward-volume, and surface spin waves without the demagnetizing field. (d)-(f) Corresponding results with the demagnetizing field included, showing enhanced comb performance due to dipole-dipole interactions, particularly for the backward-volume wave.}
    \label{fig:SM_fig3}
\end{figure*}

In addition to the COMSOL simulations, we also performed micromagnetic simulations using the Mumax3 package to validate our theoretical predictions\cite{vansteenkiste2014design}. In the Mumax3 simulations, we constructed a grid of size $256 \times 64 \times 1$ with cell dimensions of $9.18 \times 16.69 \times 20$ nm$^3$, and applied periodic boundary conditions along the $y$-direction to minimize geometric effects.

For panels (a)-(c), the demagnetizing field was turned off, and spin waves were propagated along the $x$-direction. We simulated transverse linearly polarized Floquet driving for three distinct spin-wave configurations: backward-volume waves (with magnetization oriented in-plane along $x$), forward-volume waves (magnetization oriented out-of-plane along $z$), and surface waves (magnetization oriented in-plane along $y$). Panels (d)-(f) present the corresponding results with the demagnetizing field included for the same three configurations.

In the absence of the demagnetizing field, all three spin-wave configurations exhibit identical behavior, as they can all be mapped onto an effective one-dimensional spin-chain model. When the demagnetizing field is included, the resulting frequency combs are generally enhanced compared with the demagnetization-free case, attributable to the additional dynamic nonlinearities introduced by the dipole-dipole interaction. Among the three configurations, the backward-volume wave shows the most pronounced enhancement. This can be understood from the fact that, in our model, the demagnetizing factor along the $y$-direction is the smallest. For the backward-volume wave, the transverse linearly polarized drive is applied along the $y$-direction, and the strong demagnetizing field along the $z$-direction allows the magnetization to follow the oscillating drive most effectively in the $y$-direction. In the other configurations, due to the nature of the LLG equation, the overall slow motion of the magnetization does not strictly follow the linear polarization, but instead traces a flattened elliptical trajectory.

\clearpage
\twocolumngrid        

%


\begin{thebibliography}{54}%
\makeatletter
\providecommand \@ifxundefined [1]{%
 \@ifx{#1\undefined}
}%
\providecommand \@ifnum [1]{%
 \ifnum #1\expandafter \@firstoftwo
 \else \expandafter \@secondoftwo
 \fi
}%
\providecommand \@ifx [1]{%
 \ifx #1\expandafter \@firstoftwo
 \else \expandafter \@secondoftwo
 \fi
}%
\providecommand \natexlab [1]{#1}%
\providecommand \enquote  [1]{``#1''}%
\providecommand \bibnamefont  [1]{#1}%
\providecommand \bibfnamefont [1]{#1}%
\providecommand \citenamefont [1]{#1}%
\providecommand \href@noop [0]{\@secondoftwo}%
\providecommand \href [0]{\begingroup \@sanitize@url \@href}%
\providecommand \@href[1]{\@@startlink{#1}\@@href}%
\providecommand \@@href[1]{\endgroup#1\@@endlink}%
\providecommand \@sanitize@url [0]{\catcode `\\12\catcode `\$12\catcode
  `\&12\catcode `\#12\catcode `\^12\catcode `\_12\catcode `\%12\relax}%
\providecommand \@@startlink[1]{}%
\providecommand \@@endlink[0]{}%
\providecommand \url  [0]{\begingroup\@sanitize@url \@url }%
\providecommand \@url [1]{\endgroup\@href {#1}{\urlprefix }}%
\providecommand \urlprefix  [0]{URL }%
\providecommand \Eprint [0]{\href }%
\providecommand \doibase [0]{https://doi.org/}%
\providecommand \selectlanguage [0]{\@gobble}%
\providecommand \bibinfo  [0]{\@secondoftwo}%
\providecommand \bibfield  [0]{\@secondoftwo}%
\providecommand \translation [1]{[#1]}%
\providecommand \BibitemOpen [0]{}%
\providecommand \bibitemStop [0]{}%
\providecommand \bibitemNoStop [0]{.\EOS\space}%
\providecommand \EOS [0]{\spacefactor3000\relax}%
\providecommand \BibitemShut  [1]{\csname bibitem#1\endcsname}%
\let\auto@bib@innerbib\@empty
\bibitem [{\citenamefont {Zhang}\ \emph {et~al.}(2014)\citenamefont {Zhang},
  \citenamefont {Zou}, \citenamefont {Jiang},\ and\ \citenamefont
  {Tang}}]{PhysRevLett.113.156401}%
  \BibitemOpen
  \bibfield  {author} {\bibinfo {author} {\bibfnamefont {X.}~\bibnamefont
  {Zhang}}, \bibinfo {author} {\bibfnamefont {C.-L.}\ \bibnamefont {Zou}},
  \bibinfo {author} {\bibfnamefont {L.}~\bibnamefont {Jiang}},\ and\ \bibinfo
  {author} {\bibfnamefont {H.~X.}\ \bibnamefont {Tang}},\ }\bibfield  {title}
  {\bibinfo {title} {Strongly coupled magnons and cavity microwave photons},\
  }\href {https://doi.org/10.1103/PhysRevLett.113.156401} {\bibfield  {journal}
  {\bibinfo  {journal} {Phys. Rev. Lett.}\ }\textbf {\bibinfo {volume} {113}},\
  \bibinfo {pages} {156401} (\bibinfo {year} {2014})}\BibitemShut {NoStop}%
\bibitem [{\citenamefont {Shen}\ \emph {et~al.}(2025)\citenamefont {Shen},
  \citenamefont {Li}, \citenamefont {Sun}, \citenamefont {Wu}, \citenamefont
  {Zuo}, \citenamefont {Wang}, \citenamefont {Zhu},\ and\ \citenamefont
  {You}}]{shen2025cavity}%
  \BibitemOpen
  \bibfield  {author} {\bibinfo {author} {\bibfnamefont {R.-C.}\ \bibnamefont
  {Shen}}, \bibinfo {author} {\bibfnamefont {J.}~\bibnamefont {Li}}, \bibinfo
  {author} {\bibfnamefont {Y.-M.}\ \bibnamefont {Sun}}, \bibinfo {author}
  {\bibfnamefont {W.-J.}\ \bibnamefont {Wu}}, \bibinfo {author} {\bibfnamefont
  {X.}~\bibnamefont {Zuo}}, \bibinfo {author} {\bibfnamefont {Y.-P.}\
  \bibnamefont {Wang}}, \bibinfo {author} {\bibfnamefont {S.-Y.}\ \bibnamefont
  {Zhu}},\ and\ \bibinfo {author} {\bibfnamefont {J.}~\bibnamefont {You}},\
  }\bibfield  {title} {\bibinfo {title} {Cavity-magnon polaritons strongly
  coupled to phonons},\ }\href@noop {} {\bibfield  {journal} {\bibinfo
  {journal} {Nature Communications}\ }\textbf {\bibinfo {volume} {16}},\
  \bibinfo {pages} {5652} (\bibinfo {year} {2025})}\BibitemShut {NoStop}%
\bibitem [{\citenamefont {Yu}\ \emph {et~al.}(2026)\citenamefont {Yu},
  \citenamefont {Zhou}, \citenamefont {Bauer},\ and\ \citenamefont
  {Bobkova}}]{yu2026electromagnetic}%
  \BibitemOpen
  \bibfield  {author} {\bibinfo {author} {\bibfnamefont {T.}~\bibnamefont
  {Yu}}, \bibinfo {author} {\bibfnamefont {X.-H.}\ \bibnamefont {Zhou}},
  \bibinfo {author} {\bibfnamefont {G.~E.}\ \bibnamefont {Bauer}},\ and\
  \bibinfo {author} {\bibfnamefont {I.}~\bibnamefont {Bobkova}},\ }\bibfield
  {title} {\bibinfo {title} {Electromagnetic proximity effects at
  heterointerfaces},\ }\href@noop {} {\bibfield  {journal} {\bibinfo  {journal}
  {Physics Reports}\ }\textbf {\bibinfo {volume} {1151}},\ \bibinfo {pages} {1}
  (\bibinfo {year} {2026})}\BibitemShut {NoStop}%
\bibitem [{\citenamefont {Yuan}\ \emph {et~al.}(2022)\citenamefont {Yuan},
  \citenamefont {Cao}, \citenamefont {Kamra}, \citenamefont {Duine},\ and\
  \citenamefont {Yan}}]{yuan2022quantum}%
  \BibitemOpen
  \bibfield  {author} {\bibinfo {author} {\bibfnamefont {H.}~\bibnamefont
  {Yuan}}, \bibinfo {author} {\bibfnamefont {Y.}~\bibnamefont {Cao}}, \bibinfo
  {author} {\bibfnamefont {A.}~\bibnamefont {Kamra}}, \bibinfo {author}
  {\bibfnamefont {R.~A.}\ \bibnamefont {Duine}},\ and\ \bibinfo {author}
  {\bibfnamefont {P.}~\bibnamefont {Yan}},\ }\bibfield  {title} {\bibinfo
  {title} {Quantum magnonics: When magnon spintronics meets quantum information
  science},\ }\href@noop {} {\bibfield  {journal} {\bibinfo  {journal} {Physics
  Reports}\ }\textbf {\bibinfo {volume} {965}},\ \bibinfo {pages} {1} (\bibinfo
  {year} {2022})}\BibitemShut {NoStop}%
\bibitem [{\citenamefont {Yu}\ \emph {et~al.}(2020)\citenamefont {Yu},
  \citenamefont {Lan},\ and\ \citenamefont {Xiao}}]{yu2020magnetic}%
  \BibitemOpen
  \bibfield  {author} {\bibinfo {author} {\bibfnamefont {W.}~\bibnamefont
  {Yu}}, \bibinfo {author} {\bibfnamefont {J.}~\bibnamefont {Lan}},\ and\
  \bibinfo {author} {\bibfnamefont {J.}~\bibnamefont {Xiao}},\ }\bibfield
  {title} {\bibinfo {title} {Magnetic logic gate based on polarized spin
  waves},\ }\href@noop {} {\bibfield  {journal} {\bibinfo  {journal} {Physical
  Review Applied}\ }\textbf {\bibinfo {volume} {13}},\ \bibinfo {pages}
  {024055} (\bibinfo {year} {2020})}\BibitemShut {NoStop}%
\bibitem [{\citenamefont {Grollier}\ \emph {et~al.}(2020)\citenamefont
  {Grollier}, \citenamefont {Querlioz}, \citenamefont {Camsari}, \citenamefont
  {Everschor-Sitte}, \citenamefont {Fukami},\ and\ \citenamefont
  {Stiles}}]{grollier2020neuromorphic}%
  \BibitemOpen
  \bibfield  {author} {\bibinfo {author} {\bibfnamefont {J.}~\bibnamefont
  {Grollier}}, \bibinfo {author} {\bibfnamefont {D.}~\bibnamefont {Querlioz}},
  \bibinfo {author} {\bibfnamefont {K.~Y.}\ \bibnamefont {Camsari}}, \bibinfo
  {author} {\bibfnamefont {K.}~\bibnamefont {Everschor-Sitte}}, \bibinfo
  {author} {\bibfnamefont {S.}~\bibnamefont {Fukami}},\ and\ \bibinfo {author}
  {\bibfnamefont {M.~D.}\ \bibnamefont {Stiles}},\ }\bibfield  {title}
  {\bibinfo {title} {Neuromorphic spintronics},\ }\href@noop {} {\bibfield
  {journal} {\bibinfo  {journal} {Nature electronics}\ }\textbf {\bibinfo
  {volume} {3}},\ \bibinfo {pages} {360} (\bibinfo {year} {2020})}\BibitemShut
  {NoStop}%
\bibitem [{\citenamefont {Zhou}\ and\ \citenamefont
  {Chen}(2021)}]{zhou2021prospect}%
  \BibitemOpen
  \bibfield  {author} {\bibinfo {author} {\bibfnamefont {J.}~\bibnamefont
  {Zhou}}\ and\ \bibinfo {author} {\bibfnamefont {J.}~\bibnamefont {Chen}},\
  }\bibfield  {title} {\bibinfo {title} {Prospect of spintronics in
  neuromorphic computing},\ }\href@noop {} {\bibfield  {journal} {\bibinfo
  {journal} {Advanced Electronic Materials}\ }\textbf {\bibinfo {volume} {7}},\
  \bibinfo {pages} {2100465} (\bibinfo {year} {2021})}\BibitemShut {NoStop}%
\bibitem [{\citenamefont {Kurebayashi}\ \emph {et~al.}(2026)\citenamefont
  {Kurebayashi}, \citenamefont {Finocchio}, \citenamefont {Everschor-Sitte},
  \citenamefont {Gartside}, \citenamefont {Taniguchi}, \citenamefont
  {Litvinenko}, \citenamefont {Kumar}, \citenamefont {{\AA}kerman},
  \citenamefont {Vasilaki}, \citenamefont {Sel{\c{c}}uk} \emph
  {et~al.}}]{kurebayashi2026metrics}%
  \BibitemOpen
  \bibfield  {author} {\bibinfo {author} {\bibfnamefont {H.}~\bibnamefont
  {Kurebayashi}}, \bibinfo {author} {\bibfnamefont {G.}~\bibnamefont
  {Finocchio}}, \bibinfo {author} {\bibfnamefont {K.}~\bibnamefont
  {Everschor-Sitte}}, \bibinfo {author} {\bibfnamefont {J.~C.}\ \bibnamefont
  {Gartside}}, \bibinfo {author} {\bibfnamefont {T.}~\bibnamefont {Taniguchi}},
  \bibinfo {author} {\bibfnamefont {A.}~\bibnamefont {Litvinenko}}, \bibinfo
  {author} {\bibfnamefont {A.}~\bibnamefont {Kumar}}, \bibinfo {author}
  {\bibfnamefont {J.}~\bibnamefont {{\AA}kerman}}, \bibinfo {author}
  {\bibfnamefont {E.}~\bibnamefont {Vasilaki}}, \bibinfo {author}
  {\bibfnamefont {K.}~\bibnamefont {Sel{\c{c}}uk}}, \emph {et~al.},\ }\bibfield
   {title} {\bibinfo {title} {Metrics for spin-based computing},\ }\href@noop
  {} {\bibfield  {journal} {\bibinfo  {journal} {Nature Reviews Physics}\ ,\
  \bibinfo {pages} {1}} (\bibinfo {year} {2026})}\BibitemShut {NoStop}%
\bibitem [{\citenamefont {Wang}\ \emph {et~al.}(2021)\citenamefont {Wang},
  \citenamefont {Yuan}, \citenamefont {Cao}, \citenamefont {Li}, \citenamefont
  {Duine},\ and\ \citenamefont {Yan}}]{PhysRevLett.127.037202}%
  \BibitemOpen
  \bibfield  {author} {\bibinfo {author} {\bibfnamefont {Z.}~\bibnamefont
  {Wang}}, \bibinfo {author} {\bibfnamefont {H.~Y.}\ \bibnamefont {Yuan}},
  \bibinfo {author} {\bibfnamefont {Y.}~\bibnamefont {Cao}}, \bibinfo {author}
  {\bibfnamefont {Z.-X.}\ \bibnamefont {Li}}, \bibinfo {author} {\bibfnamefont
  {R.~A.}\ \bibnamefont {Duine}},\ and\ \bibinfo {author} {\bibfnamefont
  {P.}~\bibnamefont {Yan}},\ }\bibfield  {title} {\bibinfo {title} {Magnonic
  frequency comb through nonlinear magnon-skyrmion scattering},\ }\href
  {https://doi.org/10.1103/PhysRevLett.127.037202} {\bibfield  {journal}
  {\bibinfo  {journal} {Phys. Rev. Lett.}\ }\textbf {\bibinfo {volume} {127}},\
  \bibinfo {pages} {037202} (\bibinfo {year} {2021})}\BibitemShut {NoStop}%
\bibitem [{\citenamefont {Gao}\ \emph {et~al.}(2026)\citenamefont {Gao},
  \citenamefont {Yu},\ and\ \citenamefont {Yan}}]{MFC_skr_prB}%
  \BibitemOpen
  \bibfield  {author} {\bibinfo {author} {\bibfnamefont {X.}~\bibnamefont
  {Gao}}, \bibinfo {author} {\bibfnamefont {Z.}~\bibnamefont {Yu}},\ and\
  \bibinfo {author} {\bibfnamefont {P.}~\bibnamefont {Yan}},\ }\bibfield
  {title} {\bibinfo {title} {Triple magnonic frequency combs from a skyrmionium
  in a synthetic antiferromagnet},\ }\href {https://doi.org/10.1103/c4x5-z27c}
  {\bibfield  {journal} {\bibinfo  {journal} {Phys. Rev. B}\ }\textbf {\bibinfo
  {volume} {113}},\ \bibinfo {pages} {134401} (\bibinfo {year}
  {2026})}\BibitemShut {NoStop}%
\bibitem [{\citenamefont {Liu}\ \emph {et~al.}(2024)\citenamefont {Liu},
  \citenamefont {Jin}, \citenamefont {Li}, \citenamefont {Zeng}, \citenamefont
  {Li}, \citenamefont {Yao}, \citenamefont {Cao}, \citenamefont {Zhang},\ and\
  \citenamefont {Yan}}]{liu2024low}%
  \BibitemOpen
  \bibfield  {author} {\bibinfo {author} {\bibfnamefont {X.}~\bibnamefont
  {Liu}}, \bibinfo {author} {\bibfnamefont {Z.}~\bibnamefont {Jin}}, \bibinfo
  {author} {\bibfnamefont {Z.}~\bibnamefont {Li}}, \bibinfo {author}
  {\bibfnamefont {Z.}~\bibnamefont {Zeng}}, \bibinfo {author} {\bibfnamefont
  {M.}~\bibnamefont {Li}}, \bibinfo {author} {\bibfnamefont {Y.}~\bibnamefont
  {Yao}}, \bibinfo {author} {\bibfnamefont {Y.}~\bibnamefont {Cao}}, \bibinfo
  {author} {\bibfnamefont {Y.}~\bibnamefont {Zhang}},\ and\ \bibinfo {author}
  {\bibfnamefont {P.}~\bibnamefont {Yan}},\ }\bibfield  {title} {\bibinfo
  {title} {Low-lying magnon frequency comb in skyrmion crystals},\ }\href@noop
  {} {\bibfield  {journal} {\bibinfo  {journal} {Physical Review B}\ }\textbf
  {\bibinfo {volume} {110}},\ \bibinfo {pages} {184413} (\bibinfo {year}
  {2024})}\BibitemShut {NoStop}%
\bibitem [{\citenamefont {Yao}\ \emph {et~al.}(2023)\citenamefont {Yao},
  \citenamefont {Jin}, \citenamefont {Wang}, \citenamefont {Zeng},\ and\
  \citenamefont {Yan}}]{MFC_skr_prB_THz}%
  \BibitemOpen
  \bibfield  {author} {\bibinfo {author} {\bibfnamefont {X.}~\bibnamefont
  {Yao}}, \bibinfo {author} {\bibfnamefont {Z.}~\bibnamefont {Jin}}, \bibinfo
  {author} {\bibfnamefont {Z.}~\bibnamefont {Wang}}, \bibinfo {author}
  {\bibfnamefont {Z.}~\bibnamefont {Zeng}},\ and\ \bibinfo {author}
  {\bibfnamefont {P.}~\bibnamefont {Yan}},\ }\bibfield  {title} {\bibinfo
  {title} {Terahertz magnon frequency comb},\ }\href
  {https://doi.org/10.1103/PhysRevB.108.134427} {\bibfield  {journal} {\bibinfo
   {journal} {Phys. Rev. B}\ }\textbf {\bibinfo {volume} {108}},\ \bibinfo
  {pages} {134427} (\bibinfo {year} {2023})}\BibitemShut {NoStop}%
\bibitem [{\citenamefont {Liang}\ \emph {et~al.}(2024)\citenamefont {Liang},
  \citenamefont {Cao}, \citenamefont {Yan},\ and\ \citenamefont
  {Zhou}}]{Liang2024Asymmetric}%
  \BibitemOpen
  \bibfield  {author} {\bibinfo {author} {\bibfnamefont {X.}~\bibnamefont
  {Liang}}, \bibinfo {author} {\bibfnamefont {Y.}~\bibnamefont {Cao}}, \bibinfo
  {author} {\bibfnamefont {P.}~\bibnamefont {Yan}},\ and\ \bibinfo {author}
  {\bibfnamefont {Y.}~\bibnamefont {Zhou}},\ }\bibfield  {title} {\bibinfo
  {title} {Asymmetric magnon frequency comb},\ }\href
  {https://doi.org/10.1021/acs.nanolett.4c01423} {\bibfield  {journal}
  {\bibinfo  {journal} {Nano Letters}\ }\textbf {\bibinfo {volume} {24}},\
  \bibinfo {pages} {6730} (\bibinfo {year} {2024})}\BibitemShut {NoStop}%
\bibitem [{\citenamefont {Zhou}\ \emph {et~al.}(2021)\citenamefont {Zhou},
  \citenamefont {Wang}, \citenamefont {Nie}, \citenamefont {Xia},\ and\
  \citenamefont {Guo}}]{zhou2021spin}%
  \BibitemOpen
  \bibfield  {author} {\bibinfo {author} {\bibfnamefont {Z.-w.}\ \bibnamefont
  {Zhou}}, \bibinfo {author} {\bibfnamefont {X.-g.}\ \bibnamefont {Wang}},
  \bibinfo {author} {\bibfnamefont {Y.-z.}\ \bibnamefont {Nie}}, \bibinfo
  {author} {\bibfnamefont {Q.-l.}\ \bibnamefont {Xia}},\ and\ \bibinfo {author}
  {\bibfnamefont {G.-h.}\ \bibnamefont {Guo}},\ }\bibfield  {title} {\bibinfo
  {title} {Spin wave frequency comb generated through interaction between
  propagating spin wave and oscillating domain wall},\ }\href@noop {}
  {\bibfield  {journal} {\bibinfo  {journal} {Journal of Magnetism and Magnetic
  Materials}\ }\textbf {\bibinfo {volume} {534}},\ \bibinfo {pages} {168046}
  (\bibinfo {year} {2021})}\BibitemShut {NoStop}%
\bibitem [{\citenamefont {Hou}\ \emph {et~al.}(2026)\citenamefont {Hou},
  \citenamefont {Hu},\ and\ \citenamefont {You}}]{Nonlinear_Doppler}%
  \BibitemOpen
  \bibfield  {author} {\bibinfo {author} {\bibfnamefont {J.}~\bibnamefont
  {Hou}}, \bibinfo {author} {\bibfnamefont {S.}~\bibnamefont {Hu}},\ and\
  \bibinfo {author} {\bibfnamefont {L.}~\bibnamefont {You}},\ }\bibfield
  {title} {\bibinfo {title} {Nonlinear spin-wave doppler effect for flexible
  tuning of magnonic frequencies},\ }\href {https://doi.org/10.1103/54pn-9qss}
  {\bibfield  {journal} {\bibinfo  {journal} {Phys. Rev. B}\ }\textbf {\bibinfo
  {volume} {113}},\ \bibinfo {pages} {L180411} (\bibinfo {year}
  {2026})}\BibitemShut {NoStop}%
\bibitem [{\citenamefont {Zhao}\ \emph {et~al.}(2026)\citenamefont {Zhao},
  \citenamefont {Zheng},\ and\ \citenamefont {Yan}}]{Curvature_yan}%
  \BibitemOpen
  \bibfield  {author} {\bibinfo {author} {\bibfnamefont {H.}~\bibnamefont
  {Zhao}}, \bibinfo {author} {\bibfnamefont {Q.}~\bibnamefont {Zheng}},\ and\
  \bibinfo {author} {\bibfnamefont {P.}~\bibnamefont {Yan}},\ }\bibfield
  {title} {\bibinfo {title} {Curvature-induced magnon frequency combs},\ }\href
  {https://doi.org/10.1103/v6pf-rgv8} {\bibfield  {journal} {\bibinfo
  {journal} {Phys. Rev. Lett.}\ }\textbf {\bibinfo {volume} {136}},\ \bibinfo
  {pages} {256708} (\bibinfo {year} {2026})}\BibitemShut {NoStop}%
\bibitem [{\citenamefont {Yan}\ \emph {et~al.}(2026)\citenamefont {Yan},
  \citenamefont {Bi}, \citenamefont {Wang}, \citenamefont {Flajšman},
  \citenamefont {van Dijken},\ and\ \citenamefont {Qin}}]{yan_-chip_2026}%
  \BibitemOpen
  \bibfield  {author} {\bibinfo {author} {\bibfnamefont {W.}~\bibnamefont
  {Yan}}, \bibinfo {author} {\bibfnamefont {J.}~\bibnamefont {Bi}}, \bibinfo
  {author} {\bibfnamefont {Y.}~\bibnamefont {Wang}}, \bibinfo {author}
  {\bibfnamefont {L.}~\bibnamefont {Flajšman}}, \bibinfo {author}
  {\bibfnamefont {S.}~\bibnamefont {van Dijken}},\ and\ \bibinfo {author}
  {\bibfnamefont {H.}~\bibnamefont {Qin}},\ }\bibfield  {title} {\bibinfo
  {title} {On-chip broadband magnonic frequency combs based on multi-tone
  excitation},\ }\href {https://doi.org/10.1038/s41928-026-01686-1} {\bibfield
  {journal} {\bibinfo  {journal} {Nature Electronics}\ ,\ \bibinfo {pages} {1}}
  (\bibinfo {year} {2026})}\BibitemShut {NoStop}%
\bibitem [{\citenamefont {Heins}\ \emph {et~al.}(2026)\citenamefont {Heins},
  \citenamefont {K{\"o}rber}, \citenamefont {Kim}, \citenamefont {Devolder},
  \citenamefont {Mentink}, \citenamefont {K{\'a}kay}, \citenamefont
  {Fassbender}, \citenamefont {Schultheiss},\ and\ \citenamefont
  {Schultheiss}}]{heins2026self}%
  \BibitemOpen
  \bibfield  {author} {\bibinfo {author} {\bibfnamefont {C.}~\bibnamefont
  {Heins}}, \bibinfo {author} {\bibfnamefont {L.}~\bibnamefont {K{\"o}rber}},
  \bibinfo {author} {\bibfnamefont {J.-V.}\ \bibnamefont {Kim}}, \bibinfo
  {author} {\bibfnamefont {T.}~\bibnamefont {Devolder}}, \bibinfo {author}
  {\bibfnamefont {J.~H.}\ \bibnamefont {Mentink}}, \bibinfo {author}
  {\bibfnamefont {A.}~\bibnamefont {K{\'a}kay}}, \bibinfo {author}
  {\bibfnamefont {J.}~\bibnamefont {Fassbender}}, \bibinfo {author}
  {\bibfnamefont {K.}~\bibnamefont {Schultheiss}},\ and\ \bibinfo {author}
  {\bibfnamefont {H.}~\bibnamefont {Schultheiss}},\ }\bibfield  {title}
  {\bibinfo {title} {Self-induced floquet magnons in magnetic vortices},\
  }\href@noop {} {\bibfield  {journal} {\bibinfo  {journal} {Science}\ }\textbf
  {\bibinfo {volume} {391}},\ \bibinfo {pages} {190} (\bibinfo {year}
  {2026})}\BibitemShut {NoStop}%
\bibitem [{\citenamefont {Wang}\ \emph {et~al.}(2022)\citenamefont {Wang},
  \citenamefont {Yuan}, \citenamefont {Cao},\ and\ \citenamefont
  {Yan}}]{wang2022twisted}%
  \BibitemOpen
  \bibfield  {author} {\bibinfo {author} {\bibfnamefont {Z.}~\bibnamefont
  {Wang}}, \bibinfo {author} {\bibfnamefont {H.}~\bibnamefont {Yuan}}, \bibinfo
  {author} {\bibfnamefont {Y.}~\bibnamefont {Cao}},\ and\ \bibinfo {author}
  {\bibfnamefont {P.}~\bibnamefont {Yan}},\ }\bibfield  {title} {\bibinfo
  {title} {Twisted magnon frequency comb and penrose superradiance},\
  }\href@noop {} {\bibfield  {journal} {\bibinfo  {journal} {Physical Review
  Letters}\ }\textbf {\bibinfo {volume} {129}},\ \bibinfo {pages} {107203}
  (\bibinfo {year} {2022})}\BibitemShut {NoStop}%
\bibitem [{\citenamefont {Wang}\ \emph {et~al.}(2024)\citenamefont {Wang},
  \citenamefont {Rao}, \citenamefont {Chen}, \citenamefont {Zhao},
  \citenamefont {Sun}, \citenamefont {Yao}, \citenamefont {Yu}, \citenamefont
  {Wang},\ and\ \citenamefont {Lu}}]{wang2024enhancement}%
  \BibitemOpen
  \bibfield  {author} {\bibinfo {author} {\bibfnamefont {C.}~\bibnamefont
  {Wang}}, \bibinfo {author} {\bibfnamefont {J.}~\bibnamefont {Rao}}, \bibinfo
  {author} {\bibfnamefont {Z.}~\bibnamefont {Chen}}, \bibinfo {author}
  {\bibfnamefont {K.}~\bibnamefont {Zhao}}, \bibinfo {author} {\bibfnamefont
  {L.}~\bibnamefont {Sun}}, \bibinfo {author} {\bibfnamefont {B.}~\bibnamefont
  {Yao}}, \bibinfo {author} {\bibfnamefont {T.}~\bibnamefont {Yu}}, \bibinfo
  {author} {\bibfnamefont {Y.-P.}\ \bibnamefont {Wang}},\ and\ \bibinfo
  {author} {\bibfnamefont {W.}~\bibnamefont {Lu}},\ }\bibfield  {title}
  {\bibinfo {title} {Enhancement of magnonic frequency combs by exceptional
  points},\ }\href@noop {} {\bibfield  {journal} {\bibinfo  {journal} {Nature
  Physics}\ }\textbf {\bibinfo {volume} {20}},\ \bibinfo {pages} {1139}
  (\bibinfo {year} {2024})}\BibitemShut {NoStop}%
\bibitem [{\citenamefont {Xu}\ \emph {et~al.}(2020)\citenamefont {Xu},
  \citenamefont {Zhong}, \citenamefont {Han}, \citenamefont {Jin},
  \citenamefont {Jiang},\ and\ \citenamefont {Zhang}}]{PhysRevLett.125.237201}%
  \BibitemOpen
  \bibfield  {author} {\bibinfo {author} {\bibfnamefont {J.}~\bibnamefont
  {Xu}}, \bibinfo {author} {\bibfnamefont {C.}~\bibnamefont {Zhong}}, \bibinfo
  {author} {\bibfnamefont {X.}~\bibnamefont {Han}}, \bibinfo {author}
  {\bibfnamefont {D.}~\bibnamefont {Jin}}, \bibinfo {author} {\bibfnamefont
  {L.}~\bibnamefont {Jiang}},\ and\ \bibinfo {author} {\bibfnamefont
  {X.}~\bibnamefont {Zhang}},\ }\bibfield  {title} {\bibinfo {title} {Floquet
  cavity electromagnonics},\ }\href
  {https://doi.org/10.1103/PhysRevLett.125.237201} {\bibfield  {journal}
  {\bibinfo  {journal} {Phys. Rev. Lett.}\ }\textbf {\bibinfo {volume} {125}},\
  \bibinfo {pages} {237201} (\bibinfo {year} {2020})}\BibitemShut {NoStop}%
\bibitem [{\citenamefont {Liu}\ \emph {et~al.}(2025)\citenamefont {Liu},
  \citenamefont {Zhang}, \citenamefont {Zhao}, \citenamefont {Fei},
  \citenamefont {Liu}, \citenamefont {Jia}, \citenamefont {Liu},\ and\
  \citenamefont {Xie}}]{Liu2025}%
  \BibitemOpen
  \bibfield  {author} {\bibinfo {author} {\bibfnamefont {Y.}~\bibnamefont
  {Liu}}, \bibinfo {author} {\bibfnamefont {K.-K.}\ \bibnamefont {Zhang}},
  \bibinfo {author} {\bibfnamefont {R.-S.}\ \bibnamefont {Zhao}}, \bibinfo
  {author} {\bibfnamefont {X.}~\bibnamefont {Fei}}, \bibinfo {author}
  {\bibfnamefont {X.}~\bibnamefont {Liu}}, \bibinfo {author} {\bibfnamefont
  {Z.}~\bibnamefont {Jia}}, \bibinfo {author} {\bibfnamefont {X.}~\bibnamefont
  {Liu}},\ and\ \bibinfo {author} {\bibfnamefont {X.-T.}\ \bibnamefont {Xie}},\
  }\bibfield  {title} {\bibinfo {title} {Nonreciprocal frequency combs by
  two-photon driving in cavity magnonics},\ }\href
  {https://doi.org/10.1364/OL.562248} {\bibfield  {journal} {\bibinfo
  {journal} {Opt. Lett.}\ }\textbf {\bibinfo {volume} {50}},\ \bibinfo {pages}
  {4234} (\bibinfo {year} {2025})}\BibitemShut {NoStop}%
\bibitem [{\citenamefont {Duan}\ \emph {et~al.}(2025)\citenamefont {Duan},
  \citenamefont {Cong},\ and\ \citenamefont {Shen}}]{duan2025magnon}%
  \BibitemOpen
  \bibfield  {author} {\bibinfo {author} {\bibfnamefont {Y.}~\bibnamefont
  {Duan}}, \bibinfo {author} {\bibfnamefont {A.}~\bibnamefont {Cong}},\ and\
  \bibinfo {author} {\bibfnamefont {K.}~\bibnamefont {Shen}},\ }\bibfield
  {title} {\bibinfo {title} {Magnon hybridization in easy-axis ferrimagnets},\
  }\href@noop {} {\bibfield  {journal} {\bibinfo  {journal} {Physical Review
  Applied}\ }\textbf {\bibinfo {volume} {23}},\ \bibinfo {pages} {L051006}
  (\bibinfo {year} {2025})}\BibitemShut {NoStop}%
\bibitem [{\citenamefont {Guo}\ \emph {et~al.}(2026)\citenamefont {Guo},
  \citenamefont {Gong}, \citenamefont {Lan}, \citenamefont {Guo}, \citenamefont
  {Han}, \citenamefont {Yu}, \citenamefont {Yan},\ and\ \citenamefont
  {Wang}}]{wang2026Stimulated}%
  \BibitemOpen
  \bibfield  {author} {\bibinfo {author} {\bibfnamefont {X.}~\bibnamefont
  {Guo}}, \bibinfo {author} {\bibfnamefont {T.}~\bibnamefont {Gong}}, \bibinfo
  {author} {\bibfnamefont {G.}~\bibnamefont {Lan}}, \bibinfo {author}
  {\bibfnamefont {M.}~\bibnamefont {Guo}}, \bibinfo {author} {\bibfnamefont
  {X.}~\bibnamefont {Han}}, \bibinfo {author} {\bibfnamefont {G.}~\bibnamefont
  {Yu}}, \bibinfo {author} {\bibfnamefont {P.}~\bibnamefont {Yan}},\ and\
  \bibinfo {author} {\bibfnamefont {Q.}~\bibnamefont {Wang}},\ }\bibfield
  {title} {\bibinfo {title} {Stimulated magnonic frequency combs},\ }\href
  {https://doi.org/10.1103/9d2s-4gf7} {\bibfield  {journal} {\bibinfo
  {journal} {Phys. Rev. Lett.}\ }\textbf {\bibinfo {volume} {137}},\ \bibinfo
  {pages} {036701} (\bibinfo {year} {2026})}\BibitemShut {NoStop}%
\bibitem [{\citenamefont {Huang}\ \emph {et~al.}(2026)\citenamefont {Huang},
  \citenamefont {Xue}, \citenamefont {Wang}, \citenamefont {Lei}, \citenamefont
  {Bai}, \citenamefont {Xia}, \citenamefont {Bauer},\ and\ \citenamefont
  {Yu}}]{huang2026fractionalmagnonicfrequencycombs}%
  \BibitemOpen
  \bibfield  {author} {\bibinfo {author} {\bibfnamefont {Q.-N.}\ \bibnamefont
  {Huang}}, \bibinfo {author} {\bibfnamefont {Z.}~\bibnamefont {Xue}}, \bibinfo
  {author} {\bibfnamefont {X.}~\bibnamefont {Wang}}, \bibinfo {author}
  {\bibfnamefont {Y.}~\bibnamefont {Lei}}, \bibinfo {author} {\bibfnamefont
  {L.}~\bibnamefont {Bai}}, \bibinfo {author} {\bibfnamefont {K.}~\bibnamefont
  {Xia}}, \bibinfo {author} {\bibfnamefont {G.~E.~W.}\ \bibnamefont {Bauer}},\
  and\ \bibinfo {author} {\bibfnamefont {T.}~\bibnamefont {Yu}},\ }\href
  {https://arxiv.org/abs/2606.24673} {\bibinfo {title} {Fractional magnonic
  frequency combs}} (\bibinfo {year} {2026}),\ \Eprint
  {https://arxiv.org/abs/2606.24673} {arXiv:2606.24673 [cond-mat.mtrl-sci]}
  \BibitemShut {NoStop}%
\bibitem [{\citenamefont {Dyson}(1956{\natexlab{a}})}]{Dyson1956General}%
  \BibitemOpen
  \bibfield  {author} {\bibinfo {author} {\bibfnamefont {F.~J.}\ \bibnamefont
  {Dyson}},\ }\bibfield  {title} {\bibinfo {title} {General theory of spin-wave
  interactions},\ }\href {https://doi.org/10.1103/PhysRev.102.1217} {\bibfield
  {journal} {\bibinfo  {journal} {Phys. Rev.}\ }\textbf {\bibinfo {volume}
  {102}},\ \bibinfo {pages} {1217} (\bibinfo {year}
  {1956}{\natexlab{a}})}\BibitemShut {NoStop}%
\bibitem [{\citenamefont {Dyson}(1956{\natexlab{b}})}]{Dyson1956Thermodynamic}%
  \BibitemOpen
  \bibfield  {author} {\bibinfo {author} {\bibfnamefont {F.~J.}\ \bibnamefont
  {Dyson}},\ }\bibfield  {title} {\bibinfo {title} {Thermodynamic behavior of
  an ideal ferromagnet},\ }\href {https://doi.org/10.1103/PhysRev.102.1230}
  {\bibfield  {journal} {\bibinfo  {journal} {Phys. Rev.}\ }\textbf {\bibinfo
  {volume} {102}},\ \bibinfo {pages} {1230} (\bibinfo {year}
  {1956}{\natexlab{b}})}\BibitemShut {NoStop}%
\bibitem [{\citenamefont {Holstein}\ and\ \citenamefont
  {Primakoff}(1940)}]{HolsteinPrimakoff1940}%
  \BibitemOpen
  \bibfield  {author} {\bibinfo {author} {\bibfnamefont {T.}~\bibnamefont
  {Holstein}}\ and\ \bibinfo {author} {\bibfnamefont {H.}~\bibnamefont
  {Primakoff}},\ }\bibfield  {title} {\bibinfo {title} {Field dependence of the
  intrinsic domain magnetization of a ferromagnet},\ }\href
  {https://doi.org/10.1103/PhysRev.58.1098} {\bibfield  {journal} {\bibinfo
  {journal} {Phys. Rev.}\ }\textbf {\bibinfo {volume} {58}},\ \bibinfo {pages}
  {1098} (\bibinfo {year} {1940})}\BibitemShut {NoStop}%
\bibitem [{\citenamefont {Zhang}\ \emph {et~al.}(2019)\citenamefont {Zhang},
  \citenamefont {Buscaino}, \citenamefont {Wang}, \citenamefont {Shams-Ansari},
  \citenamefont {Reimer}, \citenamefont {Zhu}, \citenamefont {Kahn},\ and\
  \citenamefont {Lon{\v{c}}ar}}]{zhang2019broadband}%
  \BibitemOpen
  \bibfield  {author} {\bibinfo {author} {\bibfnamefont {M.}~\bibnamefont
  {Zhang}}, \bibinfo {author} {\bibfnamefont {B.}~\bibnamefont {Buscaino}},
  \bibinfo {author} {\bibfnamefont {C.}~\bibnamefont {Wang}}, \bibinfo {author}
  {\bibfnamefont {A.}~\bibnamefont {Shams-Ansari}}, \bibinfo {author}
  {\bibfnamefont {C.}~\bibnamefont {Reimer}}, \bibinfo {author} {\bibfnamefont
  {R.}~\bibnamefont {Zhu}}, \bibinfo {author} {\bibfnamefont {J.~M.}\
  \bibnamefont {Kahn}},\ and\ \bibinfo {author} {\bibfnamefont
  {M.}~\bibnamefont {Lon{\v{c}}ar}},\ }\bibfield  {title} {\bibinfo {title}
  {Broadband electro-optic frequency comb generation in a lithium niobate
  microring resonator},\ }\href@noop {} {\bibfield  {journal} {\bibinfo
  {journal} {Nature}\ }\textbf {\bibinfo {volume} {568}},\ \bibinfo {pages}
  {373} (\bibinfo {year} {2019})}\BibitemShut {NoStop}%
\bibitem [{\citenamefont {Hwang}\ \emph {et~al.}(2026)\citenamefont {Hwang},
  \citenamefont {Go}, \citenamefont {Kim}, \citenamefont {Kim}, \citenamefont
  {Moon}, \citenamefont {Ju}, \citenamefont {Lee},\ and\ \citenamefont
  {Seo}}]{hwang2026electro}%
  \BibitemOpen
  \bibfield  {author} {\bibinfo {author} {\bibfnamefont {H.}~\bibnamefont
  {Hwang}}, \bibinfo {author} {\bibfnamefont {S.}~\bibnamefont {Go}}, \bibinfo
  {author} {\bibfnamefont {G.}~\bibnamefont {Kim}}, \bibinfo {author}
  {\bibfnamefont {H.-S.}\ \bibnamefont {Kim}}, \bibinfo {author} {\bibfnamefont
  {K.}~\bibnamefont {Moon}}, \bibinfo {author} {\bibfnamefont {J.~J.}\
  \bibnamefont {Ju}}, \bibinfo {author} {\bibfnamefont {H.}~\bibnamefont
  {Lee}},\ and\ \bibinfo {author} {\bibfnamefont {M.-K.}\ \bibnamefont {Seo}},\
  }\bibfield  {title} {\bibinfo {title} {Electro-optic frequency comb
  generation in lithium niobate photonic crystal fabry--p{\'e}rot
  micro-resonator},\ }\href@noop {} {\bibfield  {journal} {\bibinfo  {journal}
  {npj Nanophotonics}\ }\textbf {\bibinfo {volume} {3}},\ \bibinfo {pages} {15}
  (\bibinfo {year} {2026})}\BibitemShut {NoStop}%
\bibitem [{\citenamefont {Zhang}\ \emph
  {et~al.}(2023{\natexlab{a}})\citenamefont {Zhang}, \citenamefont {Sun},
  \citenamefont {Chen}, \citenamefont {Feng}, \citenamefont {Zhang},
  \citenamefont {Chen},\ and\ \citenamefont {Wang}}]{zhang2023power}%
  \BibitemOpen
  \bibfield  {author} {\bibinfo {author} {\bibfnamefont {K.}~\bibnamefont
  {Zhang}}, \bibinfo {author} {\bibfnamefont {W.}~\bibnamefont {Sun}}, \bibinfo
  {author} {\bibfnamefont {Y.}~\bibnamefont {Chen}}, \bibinfo {author}
  {\bibfnamefont {H.}~\bibnamefont {Feng}}, \bibinfo {author} {\bibfnamefont
  {Y.}~\bibnamefont {Zhang}}, \bibinfo {author} {\bibfnamefont
  {Z.}~\bibnamefont {Chen}},\ and\ \bibinfo {author} {\bibfnamefont
  {C.}~\bibnamefont {Wang}},\ }\bibfield  {title} {\bibinfo {title} {A
  power-efficient integrated lithium niobate electro-optic comb generator},\
  }\href@noop {} {\bibfield  {journal} {\bibinfo  {journal} {Communications
  Physics}\ }\textbf {\bibinfo {volume} {6}},\ \bibinfo {pages} {17} (\bibinfo
  {year} {2023}{\natexlab{a}})}\BibitemShut {NoStop}%
\bibitem [{\citenamefont {Hu}\ \emph {et~al.}(2022)\citenamefont {Hu},
  \citenamefont {Yu}, \citenamefont {Buscaino}, \citenamefont {Sinclair},
  \citenamefont {Zhu}, \citenamefont {Cheng}, \citenamefont {Shams-Ansari},
  \citenamefont {Shao}, \citenamefont {Zhang}, \citenamefont {Kahn} \emph
  {et~al.}}]{hu2022high}%
  \BibitemOpen
  \bibfield  {author} {\bibinfo {author} {\bibfnamefont {Y.}~\bibnamefont
  {Hu}}, \bibinfo {author} {\bibfnamefont {M.}~\bibnamefont {Yu}}, \bibinfo
  {author} {\bibfnamefont {B.}~\bibnamefont {Buscaino}}, \bibinfo {author}
  {\bibfnamefont {N.}~\bibnamefont {Sinclair}}, \bibinfo {author}
  {\bibfnamefont {D.}~\bibnamefont {Zhu}}, \bibinfo {author} {\bibfnamefont
  {R.}~\bibnamefont {Cheng}}, \bibinfo {author} {\bibfnamefont
  {A.}~\bibnamefont {Shams-Ansari}}, \bibinfo {author} {\bibfnamefont
  {L.}~\bibnamefont {Shao}}, \bibinfo {author} {\bibfnamefont {M.}~\bibnamefont
  {Zhang}}, \bibinfo {author} {\bibfnamefont {J.~M.}\ \bibnamefont {Kahn}},
  \emph {et~al.},\ }\bibfield  {title} {\bibinfo {title} {High-efficiency and
  broadband on-chip electro-optic frequency comb generators},\ }\href@noop {}
  {\bibfield  {journal} {\bibinfo  {journal} {Nature photonics}\ }\textbf
  {\bibinfo {volume} {16}},\ \bibinfo {pages} {679} (\bibinfo {year}
  {2022})}\BibitemShut {NoStop}%
\bibitem [{\citenamefont {Song}\ \emph {et~al.}(2026)\citenamefont {Song},
  \citenamefont {Lei}, \citenamefont {Xue}, \citenamefont {Cordaro},
  \citenamefont {Haas}, \citenamefont {Huang}, \citenamefont {Li},
  \citenamefont {Lu}, \citenamefont {Magalh{\~a}es}, \citenamefont {Yang} \emph
  {et~al.}}]{song2026universal}%
  \BibitemOpen
  \bibfield  {author} {\bibinfo {author} {\bibfnamefont {Y.}~\bibnamefont
  {Song}}, \bibinfo {author} {\bibfnamefont {T.}~\bibnamefont {Lei}}, \bibinfo
  {author} {\bibfnamefont {Y.}~\bibnamefont {Xue}}, \bibinfo {author}
  {\bibfnamefont {A.}~\bibnamefont {Cordaro}}, \bibinfo {author} {\bibfnamefont
  {M.}~\bibnamefont {Haas}}, \bibinfo {author} {\bibfnamefont {G.}~\bibnamefont
  {Huang}}, \bibinfo {author} {\bibfnamefont {X.}~\bibnamefont {Li}}, \bibinfo
  {author} {\bibfnamefont {S.}~\bibnamefont {Lu}}, \bibinfo {author}
  {\bibfnamefont {L.}~\bibnamefont {Magalh{\~a}es}}, \bibinfo {author}
  {\bibfnamefont {J.}~\bibnamefont {Yang}}, \emph {et~al.},\ }\bibfield
  {title} {\bibinfo {title} {Universal dynamics and microwave control of
  programmable resonant electro-optic frequency combs},\ }\href@noop {}
  {\bibfield  {journal} {\bibinfo  {journal} {Nature Physics}\ ,\ \bibinfo
  {pages} {1}} (\bibinfo {year} {2026})}\BibitemShut {NoStop}%
\bibitem [{\citenamefont {Xu}\ \emph {et~al.}(2023)\citenamefont {Xu},
  \citenamefont {Zhang}, \citenamefont {Wang}, \citenamefont {Shen},
  \citenamefont {Guo},\ and\ \citenamefont {Dong}}]{PhysRevLett.131.243601}%
  \BibitemOpen
  \bibfield  {author} {\bibinfo {author} {\bibfnamefont {G.-T.}\ \bibnamefont
  {Xu}}, \bibinfo {author} {\bibfnamefont {M.}~\bibnamefont {Zhang}}, \bibinfo
  {author} {\bibfnamefont {Y.}~\bibnamefont {Wang}}, \bibinfo {author}
  {\bibfnamefont {Z.}~\bibnamefont {Shen}}, \bibinfo {author} {\bibfnamefont
  {G.-C.}\ \bibnamefont {Guo}},\ and\ \bibinfo {author} {\bibfnamefont {C.-H.}\
  \bibnamefont {Dong}},\ }\bibfield  {title} {\bibinfo {title} {Magnonic
  frequency comb in the magnomechanical resonator},\ }\href
  {https://doi.org/10.1103/PhysRevLett.131.243601} {\bibfield  {journal}
  {\bibinfo  {journal} {Phys. Rev. Lett.}\ }\textbf {\bibinfo {volume} {131}},\
  \bibinfo {pages} {243601} (\bibinfo {year} {2023})}\BibitemShut {NoStop}%
\bibitem [{\citenamefont {Ye}\ \emph {et~al.}(2025)\citenamefont {Ye},
  \citenamefont {Sun}, \citenamefont {Zhao},\ and\ \citenamefont
  {Ma}}]{ye2025magnetostrictive}%
  \BibitemOpen
  \bibfield  {author} {\bibinfo {author} {\bibfnamefont {G.}~\bibnamefont
  {Ye}}, \bibinfo {author} {\bibfnamefont {R.}~\bibnamefont {Sun}}, \bibinfo
  {author} {\bibfnamefont {J.}~\bibnamefont {Zhao}},\ and\ \bibinfo {author}
  {\bibfnamefont {F.}~\bibnamefont {Ma}},\ }\bibfield  {title} {\bibinfo
  {title} {Magnetostrictive mechanical frequency combs},\ }\href@noop {}
  {\bibfield  {journal} {\bibinfo  {journal} {Nature Communications}\ }\textbf
  {\bibinfo {volume} {16}},\ \bibinfo {pages} {9881} (\bibinfo {year}
  {2025})}\BibitemShut {NoStop}%
\bibitem [{\citenamefont {Wang}\ \emph {et~al.}(2025)\citenamefont {Wang},
  \citenamefont {Lu}, \citenamefont {Jia},\ and\ \citenamefont
  {Xiong}}]{wang2025mechanically}%
  \BibitemOpen
  \bibfield  {author} {\bibinfo {author} {\bibfnamefont {B.}~\bibnamefont
  {Wang}}, \bibinfo {author} {\bibfnamefont {X.-H.}\ \bibnamefont {Lu}},
  \bibinfo {author} {\bibfnamefont {X.}~\bibnamefont {Jia}},\ and\ \bibinfo
  {author} {\bibfnamefont {H.}~\bibnamefont {Xiong}},\ }\bibfield  {title}
  {\bibinfo {title} {Mechanically induced magnonic frequency combs via the
  magnetostrictive effect},\ }\href@noop {} {\bibfield  {journal} {\bibinfo
  {journal} {Physical Review A}\ }\textbf {\bibinfo {volume} {111}},\ \bibinfo
  {pages} {063703} (\bibinfo {year} {2025})}\BibitemShut {NoStop}%
\bibitem [{\citenamefont {Xiong}(2023)}]{xiong2023magnonic}%
  \BibitemOpen
  \bibfield  {author} {\bibinfo {author} {\bibfnamefont {H.}~\bibnamefont
  {Xiong}},\ }\bibfield  {title} {\bibinfo {title} {Magnonic frequency combs
  based on the resonantly enhanced magnetostrictive effect},\ }\href@noop {}
  {\bibfield  {journal} {\bibinfo  {journal} {Fundamental Research}\ }\textbf
  {\bibinfo {volume} {3}},\ \bibinfo {pages} {8} (\bibinfo {year}
  {2023})}\BibitemShut {NoStop}%
\bibitem [{\citenamefont {Liu}\ \emph {et~al.}(2023)\citenamefont {Liu},
  \citenamefont {Peng},\ and\ \citenamefont {Xiong}}]{PhysRevA.107.053708}%
  \BibitemOpen
  \bibfield  {author} {\bibinfo {author} {\bibfnamefont {Z.-X.}\ \bibnamefont
  {Liu}}, \bibinfo {author} {\bibfnamefont {J.}~\bibnamefont {Peng}},\ and\
  \bibinfo {author} {\bibfnamefont {H.}~\bibnamefont {Xiong}},\ }\bibfield
  {title} {\bibinfo {title} {Generation of magnonic frequency combs via a
  two-tone microwave drive},\ }\href
  {https://doi.org/10.1103/PhysRevA.107.053708} {\bibfield  {journal} {\bibinfo
   {journal} {Phys. Rev. A}\ }\textbf {\bibinfo {volume} {107}},\ \bibinfo
  {pages} {053708} (\bibinfo {year} {2023})}\BibitemShut {NoStop}%
\bibitem [{\citenamefont {Gurevich}\ and\ \citenamefont
  {Melkov}(1996)}]{gurevich1996magnetization}%
  \BibitemOpen
  \bibfield  {author} {\bibinfo {author} {\bibfnamefont {A.~G.}\ \bibnamefont
  {Gurevich}}\ and\ \bibinfo {author} {\bibfnamefont {G.~A.}\ \bibnamefont
  {Melkov}},\ }\href@noop {} {\emph {\bibinfo {title} {Magnetization
  Oscillations and Waves}}}\ (\bibinfo  {publisher} {CRC Press},\ \bibinfo
  {address} {New York},\ \bibinfo {year} {1996})\ \bibinfo {note} {english
  translation of the Russian original (Fizmatlit, Moscow, 1994)}\BibitemShut
  {NoStop}%
\bibitem [{SM()}]{SM}%
  \BibitemOpen
  \href@noop {} {}\bibinfo {note} {See Supplemental Material at [URL] for the
  macrospin analysis of the geometric nonlinearity dictated by the unit-norm
  constraint; the classical and quantum derivations of the effective
  time-dependent two-magnon parametric modulation reduced from the
  four-particle magnon-photon process; the exact mapping of the Floquet
  Hamiltonian onto a dissipative Wannier-Stark ladder and the analytical
  derivation of the Floquet Green's function; the rigorous quantum proof of the
  angular momentum selection rules for polarized driving fields; and the
  micromagnetic simulations validating the Floquet dispersion shifts and the
  enhancement of nonlinearities by dipole-dipole interactions in thin
  films.}\BibitemShut {Stop}%
\bibitem [{\citenamefont {Rezende}(2020)}]{rezende2020fundamentals}%
  \BibitemOpen
  \bibfield  {author} {\bibinfo {author} {\bibfnamefont {S.~M.}\ \bibnamefont
  {Rezende}},\ }\href@noop {} {\emph {\bibinfo {title} {Fundamentals of
  magnonics}}},\ Vol.\ \bibinfo {volume} {969}\ (\bibinfo  {publisher}
  {Springer},\ \bibinfo {year} {2020})\BibitemShut {NoStop}%
\bibitem [{\citenamefont {Zheng}\ \emph {et~al.}(2023)\citenamefont {Zheng},
  \citenamefont {Wang}, \citenamefont {Wang}, \citenamefont {Sun},
  \citenamefont {He}, \citenamefont {Yan},\ and\ \citenamefont
  {Yuan}}]{zheng2023tutorial}%
  \BibitemOpen
  \bibfield  {author} {\bibinfo {author} {\bibfnamefont {S.}~\bibnamefont
  {Zheng}}, \bibinfo {author} {\bibfnamefont {Z.}~\bibnamefont {Wang}},
  \bibinfo {author} {\bibfnamefont {Y.}~\bibnamefont {Wang}}, \bibinfo {author}
  {\bibfnamefont {F.}~\bibnamefont {Sun}}, \bibinfo {author} {\bibfnamefont
  {Q.}~\bibnamefont {He}}, \bibinfo {author} {\bibfnamefont {P.}~\bibnamefont
  {Yan}},\ and\ \bibinfo {author} {\bibfnamefont {H.}~\bibnamefont {Yuan}},\
  }\bibfield  {title} {\bibinfo {title} {Tutorial: nonlinear magnonics},\
  }\href@noop {} {\bibfield  {journal} {\bibinfo  {journal} {Journal of Applied
  Physics}\ }\textbf {\bibinfo {volume} {134}} (\bibinfo {year}
  {2023})}\BibitemShut {NoStop}%
\bibitem [{\citenamefont {Grifoni}\ and\ \citenamefont
  {Hänggi}(1998)}]{GRIFONI1998229}%
  \BibitemOpen
  \bibfield  {author} {\bibinfo {author} {\bibfnamefont {M.}~\bibnamefont
  {Grifoni}}\ and\ \bibinfo {author} {\bibfnamefont {P.}~\bibnamefont
  {Hänggi}},\ }\bibfield  {title} {\bibinfo {title} {Driven quantum
  tunneling},\ }\href
  {https://doi.org/https://doi.org/10.1016/S0370-1573(98)00022-2} {\bibfield
  {journal} {\bibinfo  {journal} {Physics Reports}\ }\textbf {\bibinfo {volume}
  {304}},\ \bibinfo {pages} {229} (\bibinfo {year} {1998})}\BibitemShut
  {NoStop}%
\bibitem [{\citenamefont {Xu}\ \emph {et~al.}(2025)\citenamefont {Xu},
  \citenamefont {Hua}, \citenamefont {Chen},\ and\ \citenamefont
  {Yu}}]{xu_frequency_2025}%
  \BibitemOpen
  \bibfield  {author} {\bibinfo {author} {\bibfnamefont {M.}~\bibnamefont
  {Xu}}, \bibinfo {author} {\bibfnamefont {C.}~\bibnamefont {Hua}}, \bibinfo
  {author} {\bibfnamefont {Y.}~\bibnamefont {Chen}},\ and\ \bibinfo {author}
  {\bibfnamefont {W.}~\bibnamefont {Yu}},\ }\bibfield  {title} {\bibinfo
  {title} {Frequency modulation on magnons in synthetic dimensions},\ }\href
  {https://doi.org/10.1038/s41467-025-58582-z} {\bibfield  {journal} {\bibinfo
  {journal} {Nature Communications}\ }\textbf {\bibinfo {volume} {16}},\
  \bibinfo {pages} {3356} (\bibinfo {year} {2025})}\BibitemShut {NoStop}%
\bibitem [{\citenamefont {Wang}\ \emph
  {et~al.}(2026{\natexlab{a}})\citenamefont {Wang}, \citenamefont {Hua},\ and\
  \citenamefont {Yu}}]{Wang2026}%
  \BibitemOpen
  \bibfield  {author} {\bibinfo {author} {\bibfnamefont {C.}~\bibnamefont
  {Wang}}, \bibinfo {author} {\bibfnamefont {C.}~\bibnamefont {Hua}},\ and\
  \bibinfo {author} {\bibfnamefont {W.}~\bibnamefont {Yu}},\ }\bibfield
  {title} {\bibinfo {title} {Effective spin-spin coupling in magnon-polarons
  toward acoustic-assisted probing of magnetic domains},\ }\href
  {https://doi.org/10.1007/s11433-025-2884-7} {\bibfield  {journal} {\bibinfo
  {journal} {Science China Physics, Mechanics {\&} Astronomy}\ }\textbf
  {\bibinfo {volume} {69}},\ \bibinfo {pages} {247512} (\bibinfo {year}
  {2026}{\natexlab{a}})}\BibitemShut {NoStop}%
\bibitem [{com()}]{comsol}%
  \BibitemOpen
  \href {https://www.comsol.com/} {\bibinfo {title} {{COMSOL}
  {Multiphysics}{\circledR} v. 6.4. www.comsol.com. {COMSOL AB}, {Stockholm},
  {Sweden}.}}\BibitemShut {Stop}%
\bibitem [{\citenamefont {Zhang}\ \emph
  {et~al.}(2023{\natexlab{b}})\citenamefont {Zhang}, \citenamefont {Yu},
  \citenamefont {Chen},\ and\ \citenamefont {Xiao}}]{Zhang2023}%
  \BibitemOpen
  \bibfield  {author} {\bibinfo {author} {\bibfnamefont {J.}~\bibnamefont
  {Zhang}}, \bibinfo {author} {\bibfnamefont {W.}~\bibnamefont {Yu}}, \bibinfo
  {author} {\bibfnamefont {X.}~\bibnamefont {Chen}},\ and\ \bibinfo {author}
  {\bibfnamefont {J.}~\bibnamefont {Xiao}},\ }\bibfield  {title} {\bibinfo
  {title} {A frequency-domain micromagnetic simulation module based on comsol
  multiphysics},\ }\href {https://doi.org/10.1063/5.0143262} {\bibfield
  {journal} {\bibinfo  {journal} {AIP Advances}\ }\textbf {\bibinfo {volume}
  {13}},\ \bibinfo {pages} {055108} (\bibinfo {year}
  {2023}{\natexlab{b}})}\BibitemShut {NoStop}%
\bibitem [{\citenamefont {Lan}\ \emph {et~al.}(2015)\citenamefont {Lan},
  \citenamefont {Yu}, \citenamefont {Wu},\ and\ \citenamefont
  {Xiao}}]{lan_spin-wave_2015}%
  \BibitemOpen
  \bibfield  {author} {\bibinfo {author} {\bibfnamefont {J.}~\bibnamefont
  {Lan}}, \bibinfo {author} {\bibfnamefont {W.}~\bibnamefont {Yu}}, \bibinfo
  {author} {\bibfnamefont {R.}~\bibnamefont {Wu}},\ and\ \bibinfo {author}
  {\bibfnamefont {J.}~\bibnamefont {Xiao}},\ }\bibfield  {title} {\bibinfo
  {title} {Spin-{Wave} {Diode}},\ }\href
  {https://doi.org/10.1103/PhysRevX.5.041049} {\bibfield  {journal} {\bibinfo
  {journal} {Physical Review X}\ }\textbf {\bibinfo {volume} {5}},\ \bibinfo
  {pages} {041049} (\bibinfo {year} {2015})}\BibitemShut {NoStop}%
\bibitem [{\citenamefont {Wang}\ \emph
  {et~al.}(2026{\natexlab{b}})\citenamefont {Wang}, \citenamefont {Jacobs},
  \citenamefont {Englund},\ and\ \citenamefont {Trusheim}}]{wang2026ultralow}%
  \BibitemOpen
  \bibfield  {author} {\bibinfo {author} {\bibfnamefont {H.}~\bibnamefont
  {Wang}}, \bibinfo {author} {\bibfnamefont {K.}~\bibnamefont {Jacobs}},
  \bibinfo {author} {\bibfnamefont {D.~R.}\ \bibnamefont {Englund}},\ and\
  \bibinfo {author} {\bibfnamefont {M.~E.}\ \bibnamefont {Trusheim}},\
  }\bibfield  {title} {\bibinfo {title} {Ultralow-power microwave frequency
  comb at a bistable phase transition},\ }\href@noop {} {\bibfield  {journal}
  {\bibinfo  {journal} {Physical Review X}\ }\textbf {\bibinfo {volume} {16}},\
  \bibinfo {pages} {021005} (\bibinfo {year} {2026}{\natexlab{b}})}\BibitemShut
  {NoStop}%
\bibitem [{\citenamefont {Micallef}\ \emph {et~al.}(2025)\citenamefont
  {Micallef}, \citenamefont {Gu},\ and\ \citenamefont {Wu}}]{10969078}%
  \BibitemOpen
  \bibfield  {author} {\bibinfo {author} {\bibfnamefont {T.}~\bibnamefont
  {Micallef}}, \bibinfo {author} {\bibfnamefont {X.}~\bibnamefont {Gu}},\ and\
  \bibinfo {author} {\bibfnamefont {K.}~\bibnamefont {Wu}},\ }\bibfield
  {title} {\bibinfo {title} {Diode-based wideband harmonic generation for
  future nonlinear sensing applications},\ }\href
  {https://doi.org/10.1109/TMTT.2025.3557189} {\bibfield  {journal} {\bibinfo
  {journal} {IEEE Transactions on Microwave Theory and Techniques}\ }\textbf
  {\bibinfo {volume} {73}},\ \bibinfo {pages} {6180} (\bibinfo {year}
  {2025})}\BibitemShut {NoStop}%
\bibitem [{\citenamefont {Saeed}\ \emph {et~al.}(2021)\citenamefont {Saeed},
  \citenamefont {Hamed}, \citenamefont {Palacios}, \citenamefont {Uzlu},
  \citenamefont {Wang}, \citenamefont {Baskent},\ and\ \citenamefont
  {Negra}}]{9360806}%
  \BibitemOpen
  \bibfield  {author} {\bibinfo {author} {\bibfnamefont {M.}~\bibnamefont
  {Saeed}}, \bibinfo {author} {\bibfnamefont {A.}~\bibnamefont {Hamed}},
  \bibinfo {author} {\bibfnamefont {P.}~\bibnamefont {Palacios}}, \bibinfo
  {author} {\bibfnamefont {B.}~\bibnamefont {Uzlu}}, \bibinfo {author}
  {\bibfnamefont {Z.}~\bibnamefont {Wang}}, \bibinfo {author} {\bibfnamefont
  {E.}~\bibnamefont {Baskent}},\ and\ \bibinfo {author} {\bibfnamefont
  {R.}~\bibnamefont {Negra}},\ }\bibfield  {title} {\bibinfo {title}
  {Voltage-tunable thin film graphene-diode-based microwave harmonic
  generator},\ }\href {https://doi.org/10.1109/LMWC.2021.3061573} {\bibfield
  {journal} {\bibinfo  {journal} {IEEE Microwave and Wireless Components
  Letters}\ }\textbf {\bibinfo {volume} {31}},\ \bibinfo {pages} {733}
  (\bibinfo {year} {2021})}\BibitemShut {NoStop}%
\bibitem [{\citenamefont {Xiang}\ \emph {et~al.}(2023)\citenamefont {Xiang},
  \citenamefont {Zhu}, \citenamefont {Ding}, \citenamefont {Luo}, \citenamefont
  {Sun}, \citenamefont {Zeng}, \citenamefont {Zhang}, \citenamefont {Li},
  \citenamefont {Jin}, \citenamefont {Shangguan},\ and\ \citenamefont
  {Qin}}]{Xiang_2023}%
  \BibitemOpen
  \bibfield  {author} {\bibinfo {author} {\bibfnamefont {L.}~\bibnamefont
  {Xiang}}, \bibinfo {author} {\bibfnamefont {Y.}~\bibnamefont {Zhu}}, \bibinfo
  {author} {\bibfnamefont {Q.}~\bibnamefont {Ding}}, \bibinfo {author}
  {\bibfnamefont {Y.}~\bibnamefont {Luo}}, \bibinfo {author} {\bibfnamefont
  {J.}~\bibnamefont {Sun}}, \bibinfo {author} {\bibfnamefont {Z.}~\bibnamefont
  {Zeng}}, \bibinfo {author} {\bibfnamefont {J.}~\bibnamefont {Zhang}},
  \bibinfo {author} {\bibfnamefont {X.}~\bibnamefont {Li}}, \bibinfo {author}
  {\bibfnamefont {L.}~\bibnamefont {Jin}}, \bibinfo {author} {\bibfnamefont
  {Y.}~\bibnamefont {Shangguan}},\ and\ \bibinfo {author} {\bibfnamefont
  {H.}~\bibnamefont {Qin}},\ }\bibfield  {title} {\bibinfo {title} {A microwave
  comb generator based on algan/gan heterostructure schottky diodes nonlinear
  transmission line},\ }\href {https://doi.org/10.35848/1882-0786/acdcdf}
  {\bibfield  {journal} {\bibinfo  {journal} {Applied Physics Express}\
  }\textbf {\bibinfo {volume} {16}},\ \bibinfo {pages} {074001} (\bibinfo
  {year} {2023})}\BibitemShut {NoStop}%
\bibitem [{\citenamefont {Shirley}(1965)}]{shirley1965solution}%
  \BibitemOpen
  \bibfield  {author} {\bibinfo {author} {\bibfnamefont {J.~H.}\ \bibnamefont
  {Shirley}},\ }\bibfield  {title} {\bibinfo {title} {Solution of the
  schr{\"o}dinger equation with a hamiltonian periodic in time},\ }\href
  {https://doi.org/10.1103/PhysRev.138.B979} {\bibfield  {journal} {\bibinfo
  {journal} {Physical Review}\ }\textbf {\bibinfo {volume} {138}},\ \bibinfo
  {pages} {B979} (\bibinfo {year} {1965})}\BibitemShut {NoStop}%
\bibitem [{\citenamefont {Vansteenkiste}\ \emph {et~al.}(2014)\citenamefont
  {Vansteenkiste}, \citenamefont {Leliaert}, \citenamefont {Dvornik},
  \citenamefont {Helsen}, \citenamefont {Garcia-Sanchez},\ and\ \citenamefont
  {Van~Waeyenberge}}]{vansteenkiste2014design}%
  \BibitemOpen
  \bibfield  {author} {\bibinfo {author} {\bibfnamefont {A.}~\bibnamefont
  {Vansteenkiste}}, \bibinfo {author} {\bibfnamefont {J.}~\bibnamefont
  {Leliaert}}, \bibinfo {author} {\bibfnamefont {M.}~\bibnamefont {Dvornik}},
  \bibinfo {author} {\bibfnamefont {M.}~\bibnamefont {Helsen}}, \bibinfo
  {author} {\bibfnamefont {F.}~\bibnamefont {Garcia-Sanchez}},\ and\ \bibinfo
  {author} {\bibfnamefont {B.}~\bibnamefont {Van~Waeyenberge}},\ }\bibfield
  {title} {\bibinfo {title} {The design and verification of {MuMax3}},\ }\href
  {https://doi.org/10.1063/1.4899186} {\bibfield  {journal} {\bibinfo
  {journal} {AIP Advances}\ }\textbf {\bibinfo {volume} {4}},\ \bibinfo {pages}
  {107133} (\bibinfo {year} {2014})}\BibitemShut {NoStop}%
\end{thebibliography}
\end{document}